\documentclass[11pt]{article}
\usepackage{geometry}                
\usepackage{graphicx}
\usepackage{amssymb}
\usepackage{subfigure}
\usepackage{epstopdf}

\graphicspath{Diagrams}

\begin{document}

{\par\raggedleft \texttt{MAN/HEP/2007/15}\\
\texttt{19 September 2007}\par}
\bigskip{}

\bigskip{}
{\par\centering \textbf{\large Detecting Higgs bosons in the $b \bar b$ decay channel using forward proton tagging at the LHC}\large \par}
\bigskip{}

{\par\centering B.~E.~Cox, F.~K.~Loebinger and A.~D.~Pilkington \\
\par}
\bigskip{}
{\par{\small {\it School of Physics and Astronomy, University of Manchester, Manchester M13 9PL, UK}} \par}
\bigskip{}

\section*{Abstract}
A detailed study is presented of the search for Higgs bosons in the b-decay channel in the central exclusive production process at the LHC. We present results for proton tagging detectors at both 220m and 420m around ATLAS or CMS. We consider two benchmark scenarios; a Standard Model (SM) Higgs boson and the $m_h^{max}$ scenario of the Minimal Supersymmetric Standard Model (MSSM). Detector acceptance, smearing and event trigger strategies are considered. We find that the SM Higgs will be challenging to observe in the b-jet channel without improvements to the currently proposed experimental configuration, but a neutral scalar MSSM Higgs Boson could be observable in the b-jet channel with a significance of $3 \sigma$ or greater within three years of data taking at all luminosities between $2 \times 10^{33}$~cm$^{-2}$~s$^{-1}$ and $10^{34}$~cm$^{-2}$~s$^{-1}$, and at $5 \sigma$ or greater after three years in certain scenarios.
 
\newpage

\section{Introduction}

The physics potential of forward proton tagging at the LHC has attracted a great deal of attention in recent years \cite{Bialas:1991wj,Cudell:1995ki,Khoze:2001xm,Cox:2001uq,DeRoeck:2002hk,Kaidalov:2003ys,Cox:2003xp,Forshaw:2005qp,Boonekamp:2005xn,Ellis:2005fp,Bussey:2006vx,Khoze:2007td,Khoze:2007hx}. A main focus of interest is the central exclusive production (CEP) process, $pp \rightarrow p + \phi + p$, in which the protons remain intact and the central system $\phi$ is separated from the outgoing protons by a large rapidity gap. To a very good approximation, $\phi$ is constrained to be in a colour singlet, $J_z=0$, state. Observation of any particle, such as a Standard Model Higgs boson, in the central exclusive channel would therefore provide a direct observation of its quantum numbers. Furthermore, by detecting the outgoing protons and measuring their energy loss accurately, it is possible to measure the mass of the centrally produced particle regardless of its decay products \cite{Albrow:2000na}.
Because of these unique properties, it has been proposed that forward proton detectors should be installed either side of the interaction points of ATLAS and/or CMS. The FP420 collaboration has proposed to install detectors in the region 420m from the interaction points \cite{Albrow:2005ig,fp420prep}. These detectors would allow the detection of central systems in the approximate mass range 70~GeV~$< {\rm M_\phi} < 150$~GeV. Proposals also exist to upgrade the 
capabilities of ATLAS and CMS to detect protons in the 220m region \cite{Royon:2007ah,Albrow:2006xt}. These detectors, when used in conjunction  with 420m detectors, would extend the accessible mass range well beyond $150$ GeV.

\begin{figure}
\centering
\includegraphics[width=0.5\textwidth]{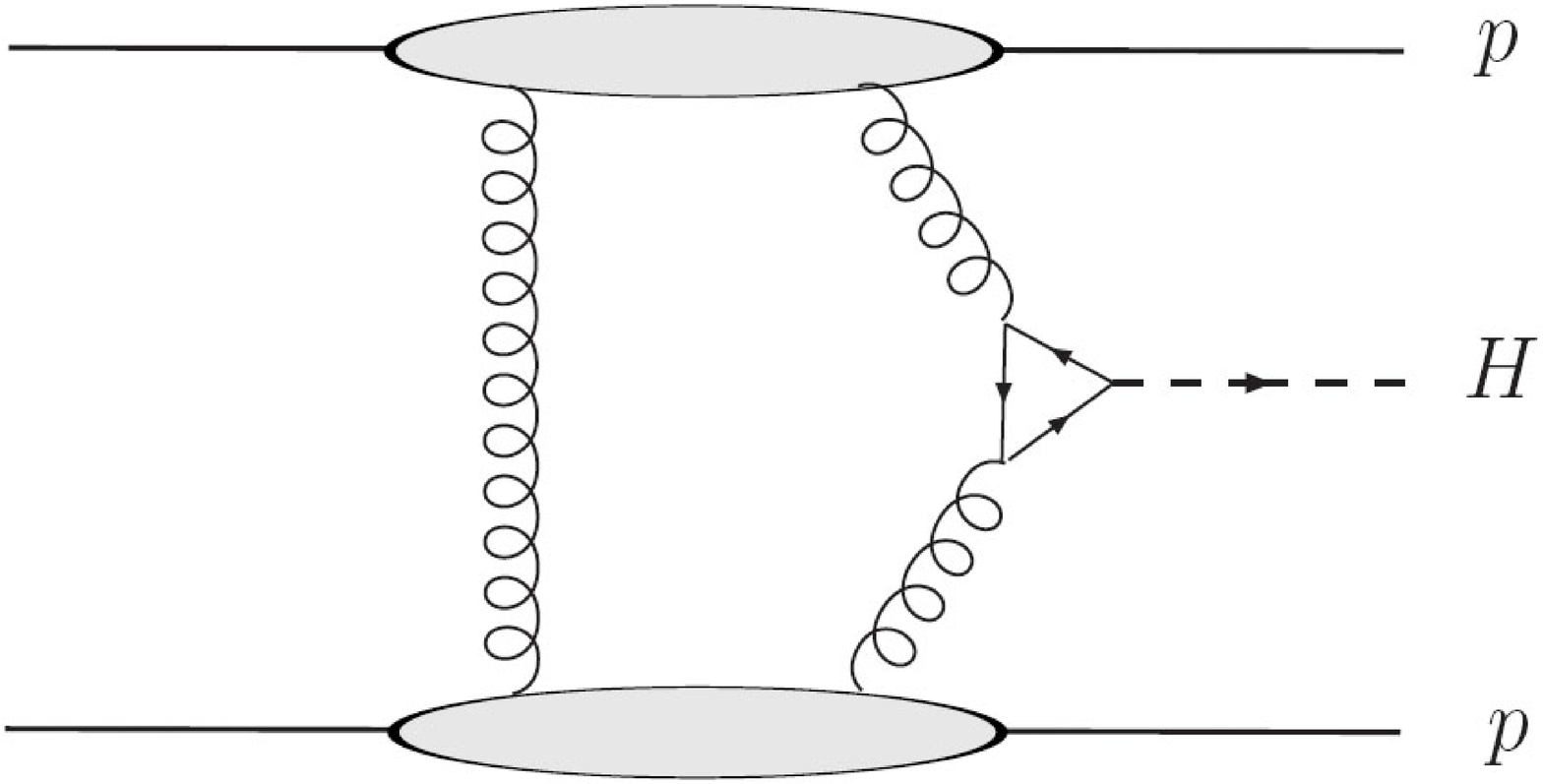}
\caption{Central exclusive production of a Higgs boson.\label{centralexclusive}}

\end{figure}

In this paper, we focus on the central exclusive production of the Standard Model (SM) Higgs boson and a Supersymmetric (MSSM) Higgs boson, with ${\rm m_h} \sim 120$ GeV. The CEP process is shown schematically in figure \ref{centralexclusive}. For this mass region, the dominant decay channel of the Higgs boson is to $b\bar{b}$, which is very difficult to observe in conventional Higgs searches at the LHC because of the large QCD background. This is not the case in central exclusive production due to the $J_z=0$ selection rule, which suppresses the leading order central exclusive $b\bar{b}$ background by a factor of $\frac{m_b^2}{M^2}$, where $M$ is the mass of the $b\bar{b}$ di-jet system. As we shall see, this renders the $b\bar{b}$ decay channel observable at the LHC in certain scenarios if appropriate proton tagging detectors are installed.

The structure of the paper is as follows. Firstly, we give a brief overview of the proposed forward detector upgrades at 220m and 420m at ATLAS and CMS, including a simulation of the acceptance of the detectors. We then discuss the predicted signal cross sections and survey the background processes. Taking a 120 GeV Standard Model Higgs boson as the benchmark, we perform a simulated analysis including an estimation of the detector acceptance and smearing effects and Level 1 trigger strategies. The analysis is then extended to the MSSM for a particular choice of parameters. 

\section{Forward detectors at the LHC}\label{fp420}

There are two properties of the proposed forward detector systems that are critical to this analysis; the acceptance of the detectors in the mass range of interest and the ability of the forward detectors to correctly associate the detected outgoing protons with a Higgs boson candidate event measured in the central detector.
This matching is critical at high luminosities, where the large number of proton-proton collisions per bunch crossing (often referred to as pile-up) leads to a high probability that forward protons from single diffractive or double pomeron (DPE) collisions not associated with a Higgs candidate event will enter the forward detectors during the same bunch crossing. The proposed forward detectors aim to associate particular protons with the Higgs candidate event by making a measurement of the outgoing proton time-of-flight (TOF) from the interaction vertex to the detectors. The difference in the arrival times of the protons on opposite sides of the central detector, $\Delta$TOF, allows a vertex measurement to be made under the assumption that the detected pair of protons originate from a single hard interaction. This vertex can then be matched with the vertex of a candidate Higgs event reconstructed using the central detector alone. The current design goal of the forward detectors is to achieve a 10ps accuracy in the TOF measurement \cite{Albrow:2005ig}, which translates into a vertex measurement accurate to 2.1mm. The use of fast-timing measurements is discussed in more detail in section \ref{olap}. It has been suggested that the central detector could also be used provide a third timing measurement \cite{White:2007ha}, which could allow for an improved rejection of pile-up events. We discuss the effect of this possibility in section \ref{sec:results}.

The acceptance of the forward detectors is governed by the distance of the active edge of the detector from the beam, which 
determines the smallest measurable energy loss of the outgoing protons, and the aperture of the LHC beam elements 
between the interaction point and the forward detectors; protons that lose too much momentum will be absorbed by beam elements, imposing an upper limit on the measurable momentum loss of the protons. 

\begin{figure}
\centering
\mbox{
	\subfigure[]{\includegraphics[width=.5\textwidth]{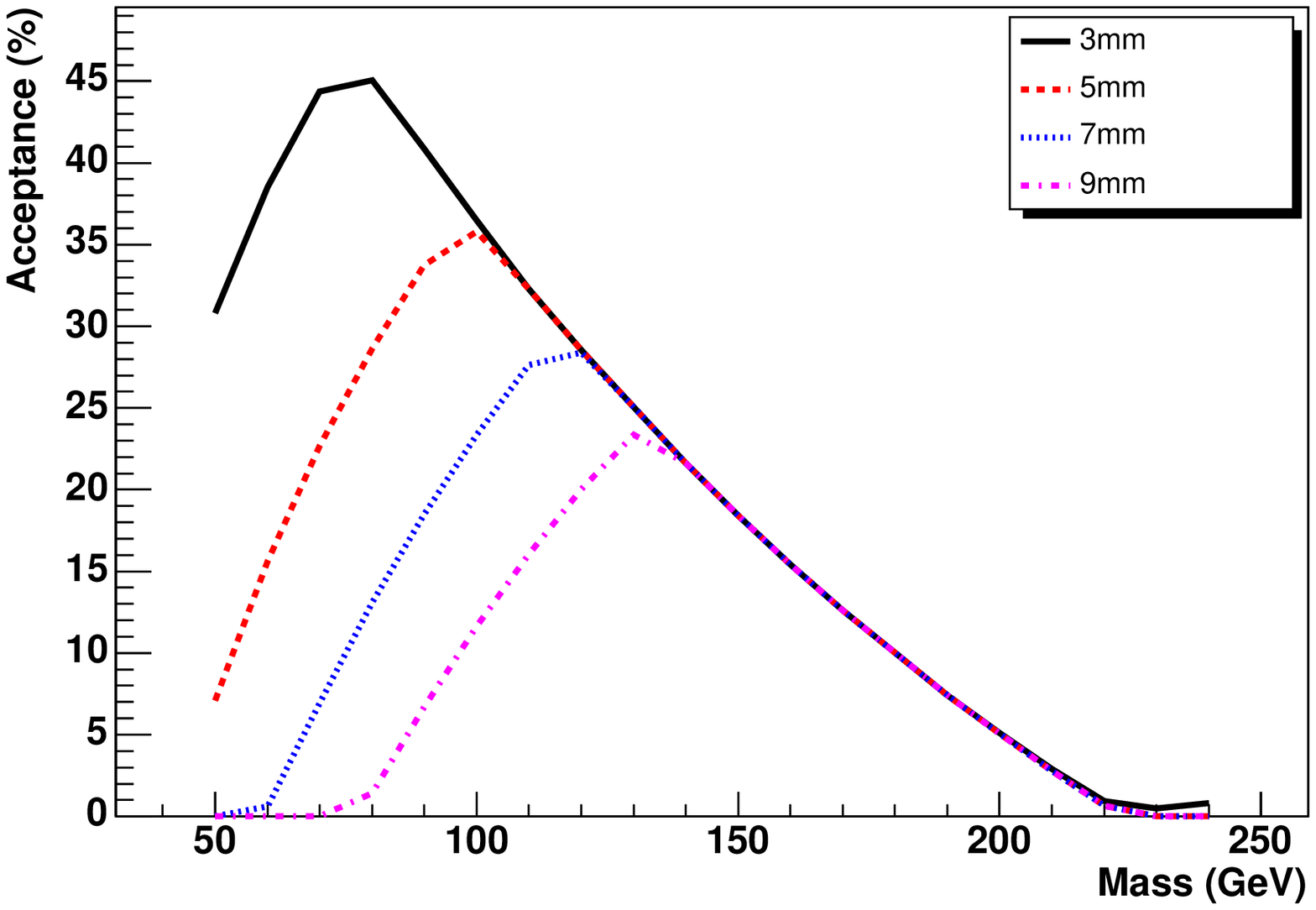}} \quad
	\subfigure[]{\includegraphics[width=.5\textwidth]{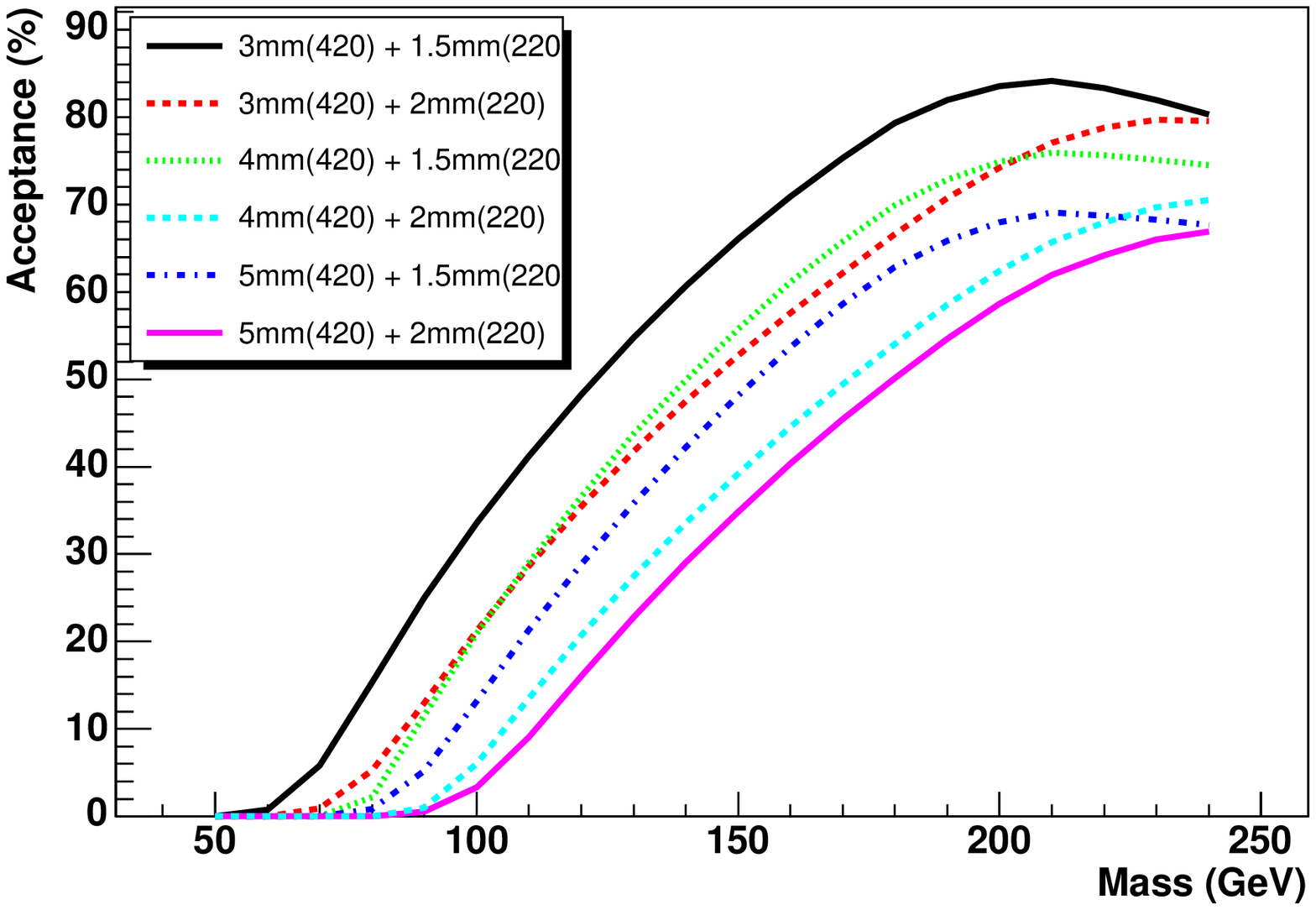}}
}
\caption{The  acceptance of the forward detectors as a function of central system mass for both protons tagged at 420m (a) or an asymmetric tag of one proton at 220m and one at 420m (b). The dependence of the acceptance on the distance of each detector from the beam is shown by the different curves.  \label{forwardaccept}}
\end{figure}

The distance of the active edge of the detector from the beam depends primarily on the beam size at each detector location. Previous estimates \cite{Alekhin:2005dyavati} have assumed that the closest distance, $D$, is given by
\begin{equation}
D \approx 10 \sigma_b + 0.5\rm{mm}
\end{equation} 
where $\sigma_{b}$ is the gaussian beam size at the detector location and 0.5 mm is a constant term that accounts for the distance from the sensitive edge of the detector to the bottom edge of the window. The beam size $\sigma_b$ is approximately 250 $\mu$m at 420m and 100 $\mu$m at 220m, leading to a distance of closest approach of 3 mm for detectors at 420m  and 1.5 mm for detectors at 220m. It is likely that the detectors will begin operation at a larger distance from the beam, at least until the detectors and machine background conditions are well understood \cite{fp420prep,appleby}. Figure \ref{forwardaccept} (a) shows the acceptance for events in which both outgoing protons are detected at 420m around IP1 (ATLAS), as a function of the mass of the central system for different detector distances from the beam. The protons are generated using the ExHuME Monte Carlo \cite{Monk:2005ji} and tracked through the LHC beam lattice using the FPTrack program with version 6.500 of the LHC optics \cite{fptrack}. The central system mass, $M$, is calculated from the forward proton momenta,
\begin{equation}\label{cepmass}
M^2 \simeq \xi_1 \, \xi_2 s,
\end{equation}
where the $\xi_i$ are the fractional momentum losses of the protons and $\sqrt{s}$ is the centre-of-mass energy of the collision. The acceptance for a 120 GeV Higgs boson is independent of the distance of approach of the detectors from the beam up to approximately 7mm in the case where both protons are detected at 420m. This is consistent with the findings of \cite{Albrow:2005ig}.

The situation is very different for events in which one proton is tagged at 220m and the other at 420m, as shown in figure \ref{forwardaccept} (b). In this case, the acceptance is increased when either detector is moved closer to the beam. For a 120 GeV Higgs boson, with 420m detectors at 5mm and 220m detectors at 2mm, the acceptance is 28\% if both protons are tagged at 420m (symmetric tag) with an additional 16\% acceptance if one proton is tagged at 220m and one at 420m (asymmetric). Moving the 420m detectors inwards to 3mm and the 220m detectors to 1.5mm increases the asymmetric acceptance by up to a factor of three, as shown in figure \ref{forwardaccept} (b). Figures \ref{forwardaccept} (a) and (b) also demonstrate the increasing importance of 220m detectors as the mass of the central system increases. At IP5 (CMS) the symmetric acceptance is identical to that at IP1. For the asymmetric tags, however, the acceptance is worse by a factor of $\sim 3$ across the mass range of interest. This is caused by the horizontal (rather than vertical) plane of the crossing angle of the beams at IP5 \cite{fp420prep}. In this paper we concentrate on the measurement around IP1 (ATLAS).     

\section{Central exclusive $H\rightarrow b\bar{b}$ production}
\label{scenarios}

\begin{figure}
\centering

\includegraphics[width=0.5\textwidth]{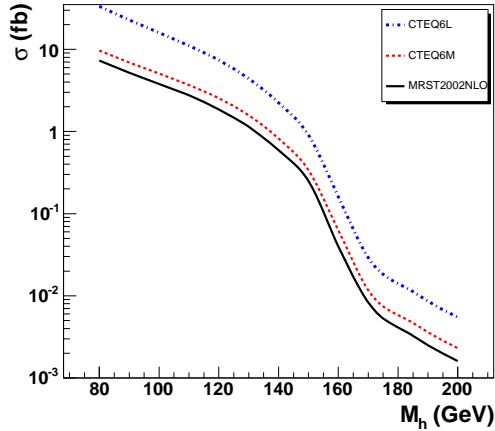}

\caption{The cross section for the central exclusive production of a Standard Model Higgs Boson decaying to b-quarks as a function of $m_h$, for three different proton parton distribution functions.\label{sigmamh}}

\end{figure}

Central exclusive signal and background events are simulated with parton showering and hadronisation effects using ExHuME v1.3.4 \cite{Monk:2005ji}. ExHuME contains a direct implementation of the Khoze, Martin and Ryskin (KMR) calculation of the central exclusive production process \cite{Khoze:2001xm,Khoze:2000cy}. 
The cross section for the CEP of a Standard Model Higgs boson decaying to $b \bar b$ as a function of the Higgs mass is shown in figure \ref{sigmamh}. The rapid decrease in cross section at $m_h\sim 160$ GeV occurs because the primary decay channel changes from $b\bar{b}$ to $WW^{(*)}$. For masses above $140$ GeV, it is expected that the Higgs boson should be observed in the $WW^{(*)}$ channel with a luminosity of 30 fb$^{-1}$ using forward proton tagging \cite{Cox:2005if}. 

The primary uncertainties in the predicted cross section come from two sources; the parton distribution functions (pdf) and the soft survival (often termed rapidity gap survival) factor. The CEP cross sections are relatively sensitive to the pdfs because the derivative of the gluon density enters to the fourth power.   
Figure \ref{sigmamh} shows the cross section prediction for three different pdf choices, CTEQ6M, MRST2002NLO and CTEQ6L. The cross section for the CEP of a Standard Model Higgs boson of mass $m_h = 120$ GeV decaying into b-quarks varies from 1.86~fb (MRST2002NLO) to 7.38~fb(CTEQ6L1). The spread of a factor of $\sim 4$ is consistent with the findings of  \cite{Forshaw:2005qp}.  For the purposes of this analysis, we chose CTEQ6M as our default pdf as it lies between the two extremes. There is some justification for choosing an NLO pdf because the KMR calculation contains a NLO QCD K-factor (1.5) for SM Higgs Boson production. 

The soft survival factor, $S^2$, is the probability that there are no additional hard scatters in a single $pp$ collision and is expected to vary from process to process. For CEP processes, we take the ExHuME default of  $S^2 = 0.03$ at LHC energies. Until very recently, there was a consensus that $S^2$ should be between $2.5\%$ and $4.0\%$ for the CEP of a Higgs boson at the LHC \cite{Alekhin:2005dx}. Two very recent studies have predicted a lower value \cite{Gotsman:2007ac,Frankfurt:2006jp,strikman} and it remains to be seen whether a new theoretical consensus can be reached. In any case, $S^2$ will be measurable in early LHC data.  

The MSSM contains three neutral Higgs bosons - two scalar and one pseudo-scalar. The $J_Z =0$, parity-even selection rules strongly suppress CEP of the pseudo-scalar. This can be advantageous in areas of MSSM parameter space where the pseudo-scalar is almost degenerate in mass to one (or both) of the scalar Higgs bosons, since CEP would provide a clean and complimentary measurement of the mass of the scalar only \cite{Kaidalov:2003fw}, and allow nearly-degenerate Higgs Bosons to be distinguished \cite{Ellis:2005fp}. Furthermore, at large values of tan$\beta$, the cross section for the CEP of the scalar Higgs bosons can be strongly enhanced relative to the CEP of a SM Higgs boson of the same mass \cite{Heinemeyer:2007tu}. We choose a point in parameter space defined by the $m_h^{\rm{max}}$ scenario \cite{Carena:2002qg} with $m_A = 120$ GeV and tan$\beta=40$, resulting in the scalar Higgs boson having a mass of 119.5 GeV and a decay width of 3.3 GeV. With this choice of parameters, the lightest scalar Higgs boson has an enhanced CEP cross section and is almost degenerate in mass with the pseudo-scalar.  The CEP cross section for the lightest Higgs Boson decaying to b-quarks in this scenario is predicted to be $20.5$ fb. The uncertainty on this prediction is the same as that for the SM Higgs boson.

\section{Backgrounds to the $H\rightarrow b \bar{b}$ channel}\label{backgrounds}

The backgrounds to central exclusive Higgs production can be broken down into three categories; central exclusive di-jet production, double pomeron exchange and overlap. 

\subsection{Central exclusive di-jet backgrounds}
The most difficult backgrounds to deal with are the central exclusive (CEP) di-jet backgrounds. Central exclusive $b\bar{b}$ production is suppressed at leading order by the $J_Z=0$ selection rule, but is still present and forms an irreducible continuum background beneath the Higgs mass peak. Central exclusive glue-glue production has a much larger cross section than $b\bar{b}$ production since it is not suppressed. It contributes to the background when the gluon jets are mis-identified as b-jets (1.3\% probability for each mis-tag at ATLAS - see section \ref{detector}). Central exclusive $c\bar{c}$ events do not contribute to the background for two reasons. Firstly, $c\bar{c}$ production is suppressed with respect to the $b\bar{b}$ process by $\frac{m_c^2}{m_b^2}$. Secondly, the $c\bar{c}$ background is further suppressed by a factor of $36$ relative to the $b\bar{b}$ cross section assuming that the probability of mis-identifying a c-quark is approximately 0.1 for a b-tag efficiency of 0.6 \cite{btag}. 

Higher order CEP di-jet backgrounds such as the 3-jet final state $b \bar b g$ process have been studied in \cite{Khoze:2006um}. It is expected that these backgrounds will be lower than the LO $gg$ and $b \bar b$ backgrounds after experimental cuts, but they should be implemented into the ExHuME Monte Carlo and a full study performed before definitive conclusions can be drawn. 

\subsection{Double Pomeron backgrounds} 
\begin{figure}
\centering
\includegraphics[width=0.5\textwidth]{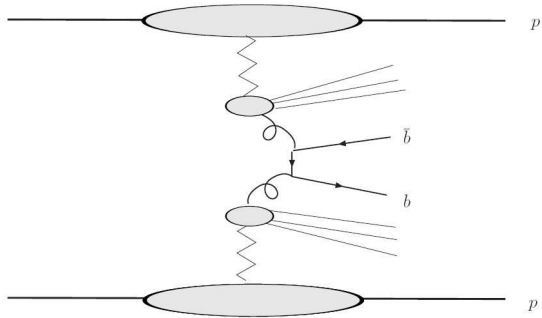}
\caption{Production of a $b\bar{b}$ system via double pomeron exchange (DPE). \label{dpepomeron}}
\end{figure}

Double pomeron exchange (DPE) is defined as the process $pp \rightarrow p + X + p$, where $X$ is a central system produced by pomeron-pomeron fusion ($\rm{I \! P}\rm{I \! P}$). The pomeron is assigned a partonic structure and so there are always pomeron remnants accompanying the hard scatter as shown in figure \ref{dpepomeron}. The relevant background processes are separated into $b\bar{b}$ and $jj$, where $j$ represents light-quark and gluons jets. 

The DPE events are simulated using the POMWIG v2.0 event generator \cite{Cox:2000jt}, which implements the diffractive parton distribution functions measured by the H1 Collaboration \cite{Aktas:2006hx,Aktas:2006hy}. We use H1 2006 fit B, although we discuss the effect of using different diffractive pdfs. We choose the POMWIG option to treat the valence partons in the pomeron as gluons. POMWIG is normalised to the H1 data, and therefore does not account for the soft survival factor. For hard double pomeron events, we take $S^2 = 0.03$. POMWIG is also capable of generating scatters involving sub-leading (non-diffractive) exchanges ($\rm{I \! R}\rm{I \! R}$). We do not include this contribution because the cross section for $\rm{I \! R}\rm{I \! R} \rightarrow$ $b\bar{b}$ events is $\sim$ 0.034 fb for $\xi_{1,2} \leq 0.05$, $80 {\rm GeV} \leq {\rm M} \leq 160$ GeV and $b$-quark $p_T \geq 35$ GeV. This is negligible compared to the cross section for $\rm{I \! P}\rm{I \! P} \rightarrow$ $b\bar{b}$, which is $\sim$65 fb in the same kinematic region.

\subsection{Overlap backgrounds}
\label{olap}
Overlap events are defined as a coincidence between an event that produces a central system of interest, $X$, and one or more diffractive events (single diffractive $pp \rightarrow pX$ or DPE) which produce protons in the acceptance range of a forward detector in the same bunch crossing. On average there will be 3.5 interactions per bunch crossing at low instantaneous luminosity (10$^{33}$ cm$^{-2}$ s$^{-1}$) and 35 interactions at high instantaneous luminosity (10$^{34}$ cm$^{-2}$ s$^{-1}$). We investigate three types of overlap event; [p][X][p], [pp][X]  and [pX][p]. The square brackets specify the interaction to which a part of the overlap event belongs - in this notation both the CEP and DPE events would be [pXp]. 

The cross section, $\sigma_{olap}$, for the overlap background may be estimated by
\begin{eqnarray} \label{olapxs}
\sigma_{olap} & = & \sigma_{[X]} \, \left[ \, \sum_{N=3}^{\infty} \frac{\lambda^{N} e^{-\lambda}}{N!} \,  P_{2[p]}\left(N-1\right) \, + \, \sum_{N=2}^{\infty}  \frac{\lambda^{N} e^{-\lambda}}{N!} \,  P_{[pp]} \left(N-1\right)  \right] \nonumber \\
& & + \, \, \sigma_{[pX]} \, \sum_{N=2}^{\infty} \frac{\lambda^{N} e^{-\lambda}}{N!} \,  P_{[p]} \left(N-1\right), 
\end{eqnarray} 
where $\sigma_{[X]}$ is the inclusive ($pp \rightarrow {\rm jets}$) di-jet cross section, $\lambda$ is the average number of $pp$ interactions per bunch crossing and $N$ is the actual number of interactions in a specific bunch crossing. Because the actual number of interactions is not fixed, we sum over all possible numbers and weight each configuration by a Poisson distribution. In the first term, $P_{2[p]}\left(n\right)$ is the probability that, given $n$ interactions, there are at least two events that produce a forward proton (one on each side of the interaction point) by any mechanism. This is dominated by soft single diffractive events $pp \rightarrow pX$. $P_{2[p]}\left(n\right)$  is given by a trinomial distribution, i.e.
\begin{equation}
P_{2[p]}\left(n\right) = \sum_{r+q=2}^{n} \sum_{q=1}^{r+q-1} \frac{n!}{\left(n-\left[r+q\right]\right)! \, r! \, q!}
\, \left( f_{[p]}^{+}\right)^{r} \, \left(f_{[p]}^{-}\right)^{q} \, \left( 1 - f_{[p]}^{+} - f_{[p]}^{-}\right) ^{n-r-q} 
\end{equation}
where, for example, $f_{[p]}^{+}$ is the fraction of events at the LHC that produce a forward proton in the $+z$ direction and within the forward detector acceptance. 

In the second term, $P_{[pp]}(n)$ is defined as the probability that there is at least one event that contains an outgoing proton on each side of the interaction point within the acceptance of the forward detectors (dominated by the soft double pomeron process $pp \rightarrow pp X$). $P_{[pp]}(n)$ is a binomial distribution that utilises the event fraction $f_{[pp]}$. Note that double pomeron events 
can also contribute to the first term in equation \ref{olapxs}, in the case where only one of the outgoing protons falls within the detector acceptance.

The third term deals with a two-fold coincidence between a single diffractive di-jet event ($pp \rightarrow pX$), which produces a proton 
within the forward detector acceptance AND a hard central diffractive system $X$ that mimics the signal, and an overlap event that produces a proton on the opposite side. $\sigma_{[pX]}$ is the total single diffractive di-jet cross section ($pp \rightarrow pX$), i.e the outgoing proton can be on either side of the interaction point. $P_{[p]}(n)$ is defined as the probability that there is at least one event with a forward proton in the forward detector acceptance on the opposite side of the IP to the single diffractive proton from the hard event. This is defined in a similar way to $P_{[pp]}(n)$, but using the event fraction $f_{[p]}$.  

\begin{figure}
\centering
\mbox{
	\subfigure[]{\includegraphics[width=.5\textwidth]{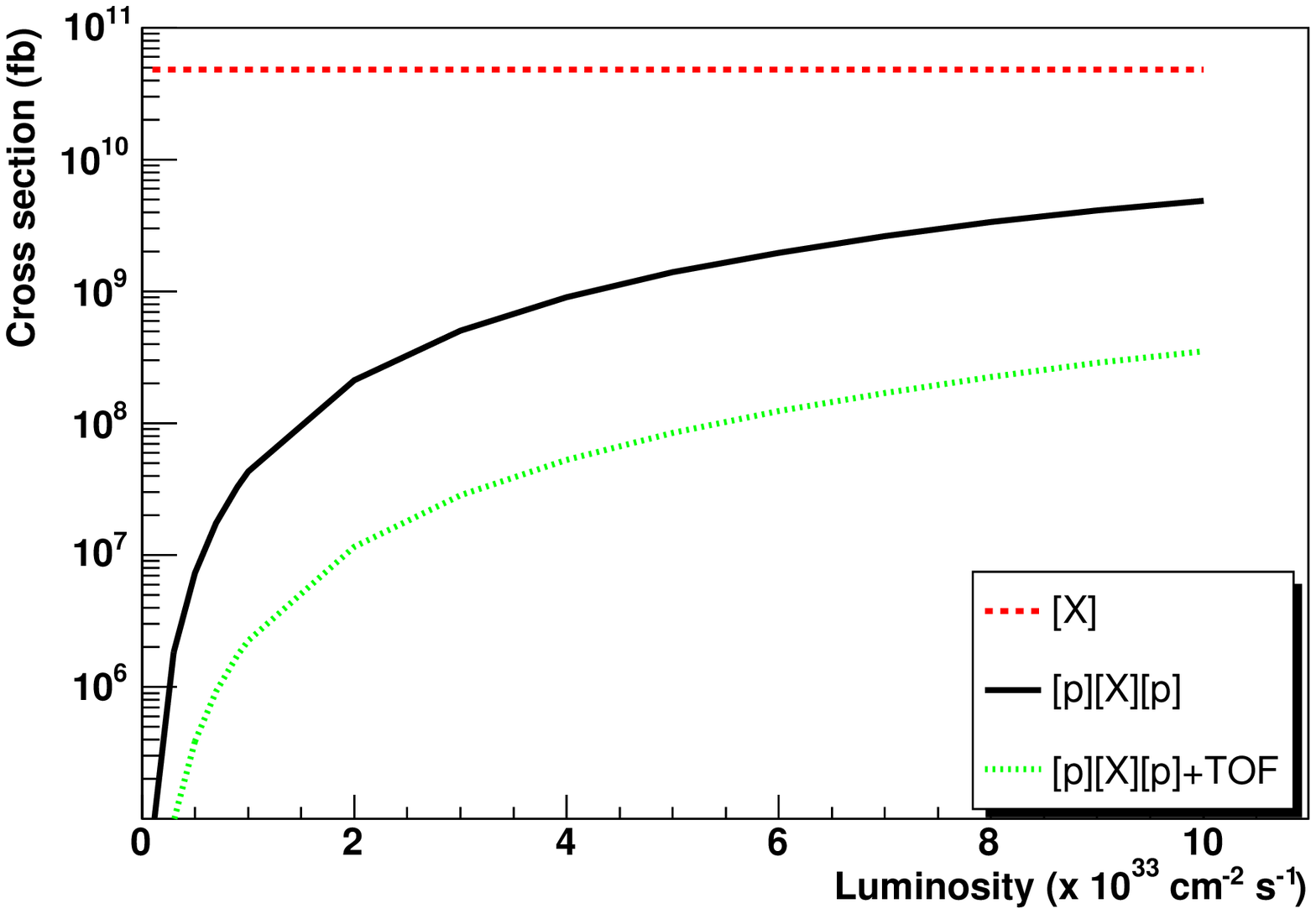}} \quad
	\subfigure[]{\includegraphics[width=.5\textwidth]{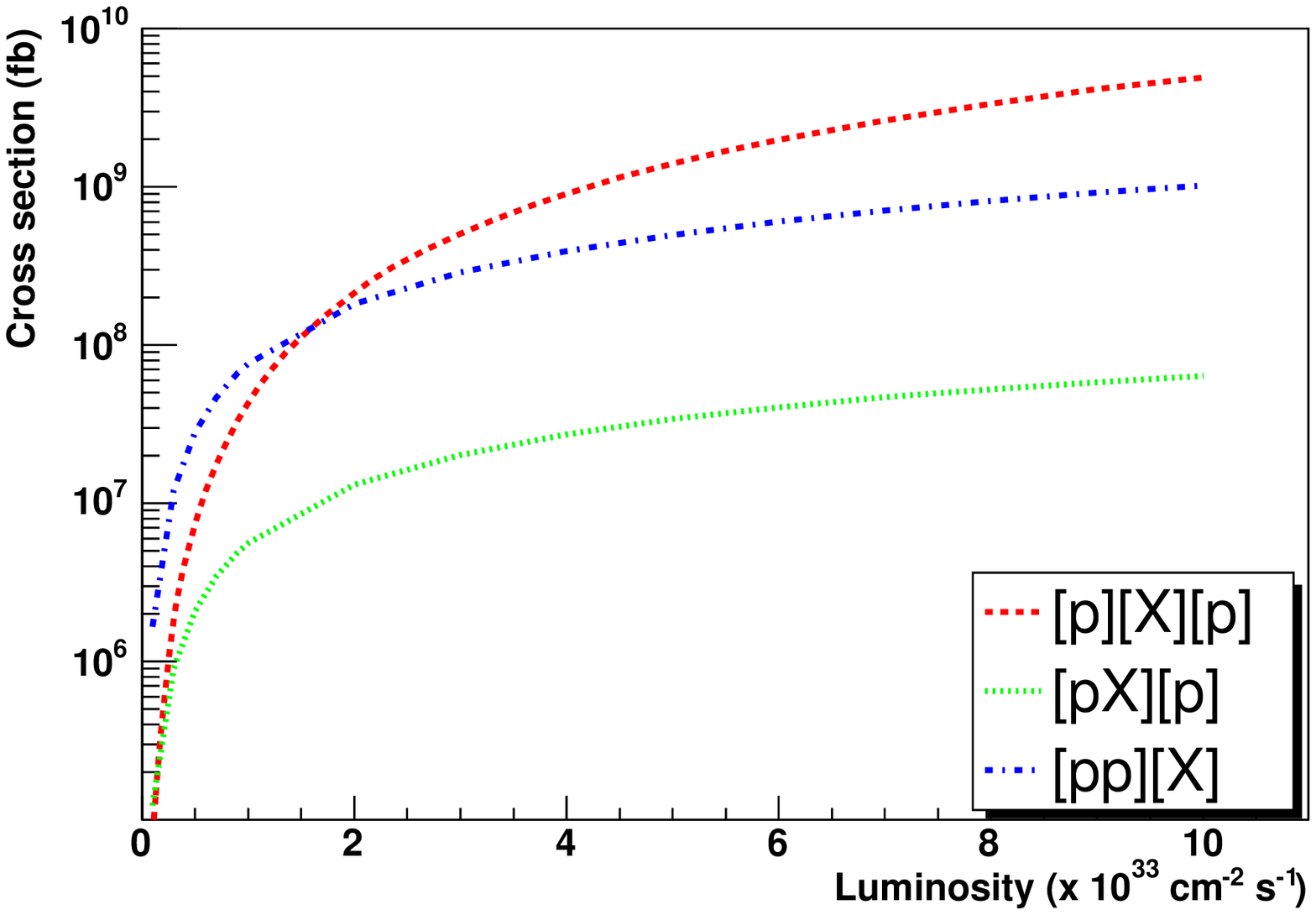}}
}
\caption{Figure (a) shows the di-jet [p][X][p] background cross section for both protons tagged at 420m. The di-jet cross section is calculated using HERWIG + JIMMY for partons with $E_T\geq40 \rm{GeV}$. Figure (b) shows the cross sections for each of the three types of overlap background. \label{olaplumidep}}
\end{figure}

For small proton momentum losses, the cross section for events at the LHC that contain a forward proton is dominated by the single diffractive cross section, $\sigma_{SD}$ \cite{Khoze:2006gg};
\begin{equation} \label{sdxs}
\frac{1}{\sigma_T}\frac{d\sigma^{SD}}{dtd\xi} = - \frac{g_N^2(t) g_{3\rm{I \! P}}}{16\pi^2 g_N(0)}
 \xi^{\alpha_{\rm{I \! P}}(0) - 2 \alpha_{\rm{I \! P}}(t)} S_{SD}^2(s,t),
\end{equation}
where $\sigma_T$ is the total cross section at the LHC, $S^2_{SD}$ is the soft-survival factor for soft single diffraction (0.087), $\alpha_{\rm{I \! P}}(t)$ is the pomeron trajectory, $t$ is the squared 4-momentum transfer at the proton vertex, $g_N(t)$ is the pomeron nucleon coupling and $g_{3\rm{I \! P}}$ is the triple pomeron vertex.
The value of $f_{[p]}$ is calculated by integrating equation \ref{sdxs} using standard Monte Carlo techniques, for a specific acceptance range in $\xi$ and $t$.
The cross section for the [p][X][p] di-jet events, for parton $p_T > 40$GeV and $0.002 \leq \xi \leq 0.02$, $-5 \leq t \leq 0$ is shown in figure \ref{olaplumidep} (a). The cross section increases by two orders of magnitude from low to high luminosity.

At larger values of $\xi$, there will be an additional contribution from non-diffractive (i.e. reggeon exchange) events. This contribution is estimated using the PYTHIA \cite{Sjostrand:2001yu} and PHOJET \cite{Engel:1995yd} event generators. The predictions for the diffractive and non-diffractive contributions to $f_{[p]}$ are compared to equation \ref{sdxs} in table \ref{fractions}. As expected, the diffractive contribution dominates at small $\xi$. The non-diffractive contribution becomes increasingly important at higher $\xi$, but remains smaller than the diffractive contribution for $\xi \leq 0.05$. PHOJET predicts a higher fraction of non-diffractive events than PYTHIA and also predicts that a large fraction of forward protons are due to DPE events. In this analysis we use the prediction for single diffraction given by equation \ref{sdxs} for $\xi < 0.02$ (which would give hits only at 420m), and take the non-diffractive contribution to be negligible. For $0.02 \le \xi \le 0.2$, we add the PYTHIA prediction for the non-diffractive component, since this agrees more closely with theoretical expectations that the non-diffractive fraction of $pp$ collisions at the LHC in this kinematic range should be between 1.0\% and 1.7\% \cite{ryskinkhoze}. 

\begin{table}[t]
\centering
\begin{tabular}{|cc|c|c|c|c|}
\hline
\multicolumn{2}{|c|}{Generator} & \multicolumn{4}{c|}{$\xi$ range} \\
 & & 0.002 - 0.02 & 0.02 - 0.05 & 0.02 - 0.1 & 0.02 - 0.2\\
 \hline
 Equation 5 & ($f_{SD}$) & 0.0112 & 0.0040 & 0.0070 & 0.0098 \\
 \hline
 PYTHIA &($f_{SD}$) & 0.0104 & 0.0045 & 0.0081 & 0.0112 \\
 & ($f_{ND}$) & 0.0002 & 0.0016 & 0.0043 & 0.0124  \\
 \hline
 PHOJET & ($f_{SD}$) & 0.0069 & 0.0031 & 0.0055 & 0.0081\\
 & ($f_{SD}$ $+$ $f_{DPE}$) & 0.0097 & 0.0045 & 0.0081 & 0.0118 \\
 & ($f_{ND}$) & 0.0018 & 0.0025 & 0.0059 & 0.0192\\
\hline
\end{tabular}

\caption{The fraction of events at the LHC that produce a forward proton on one side of the interaction point in a specific kinematic range. The PYTHIA and PHOJET event generators are compared to the single diffractive cross section given in equation \ref{sdxs}. SD labels the outgoing proton from single diffractive scatters and  ND labels the protons produced from non-diffractive scatters. DPE labels double pomeron exchange events. \label{fractions}}
\end{table}%

The fraction of soft double pomeron events (DPE) within the detector acceptance, $f_{[pp]}$, is estimated using the PHOJET event generator. PHOJET predicts that the fraction of DPE events that have both forward protons in the range $0.002 \leq \xi \leq 0.02$ is  $6.3 \times 10^{-4}$. The fraction of DPE events that have one proton with $0.002 \leq \xi \leq 0.02$ and one with $0.02 \leq \xi \leq 0.2$ is $1.4 \times 10^{-3}$. This allows us to construct the [pp][X] overlap backgrounds. 
The cross sections for the three types of di-jet overlap backgrounds, [p][X][p], [pX][p] and [pp][X] are shown in figure \ref{olaplumidep} (b) as a function of instantaneous luminosity. 

Monte Carlo overlap events used for the simulated analysis presented in section \ref{cepvars} are constructed as follows: The HERWIG Monte Carlo \cite{Corcella:2002jc} is used to generate di-jet events, interfaced to JIMMY \cite{Butterworth:1996zw} to simulate secondary scatters between spectator partons in the interacting protons. Single diffractive di-jet events are generated using POMWIG. 
For small $\xi$, diffractive protons are generated according to the distribution in $\xi$ and $t$ given by equation \ref{sdxs} and added to the event record. For $\xi > 0.02$, the protons are selected as diffractive or non-diffractive on an event-by-event basis, according to the cross sections presented in section \ref{olap}. We model the non-diffractive proton kinematics using a Regge flux factor, given by
\begin{equation}
\frac{d^2 \sigma}{d\xi dt} \,  \propto \,  \frac{e^{b_{\rm{I \! R}} t}}{\xi^{2\alpha_{\rm{I\!R}}(t)-1} }
\end{equation}
where $\alpha_{\rm{I\!R}}(t) =0.5 + 0.3t$ and $b_{\rm{I\!R}}=1.6$ GeV$^{-2}$ \cite{Aktas:2006hx,Aktas:2006hy}.

\section{Simulated measurement of central exclusive Higgs Boson production in the b-jet channel}\label{cepvars}

Having generated the signal and background events, we now investigate a possible experimental strategy for detecting central exclusive Higgs production in the $b\bar{b}$ decay channel. The analysis presented here is specific to the ATLAS detector, in that we use the ATLAS 
parameters for detector smearing effects and b-tagging efficiencies. We do not expect the conclusions to 
be materially different for CMS for 420m detectors alone. However, as discussed in section \ref{fp420} the asymmetric events, with one proton detected at  220m and one at 420m, have a lower acceptance at CMS due to the different LHC optics. All figures are presented for the symmetric tag case, i.e. 
a proton hit within the acceptance of a 420m detector on each side of the central detector \footnote{We apply a  
cut of $80 \leq \Delta M \leq 160$ GeV on the mass $M$ of the central system as measured in the forward proton detectors to restrict the size of the Monte Carlo samples before making the figures.}. 
We present numerical results for both symmetric and asymmetric (420m + 220m) tags in section \ref{sec:results}.    

\subsection{Triggering on CEP b-jet events}
\label{trigger}
Proton tagging detectors at 420m are too far away to be included in the level one (L1) trigger decision at ATLAS or CMS in normal running and with the standard trigger hardware. It will therefore be necessary to design a trigger strategy that can keep the symmetric 420m + 420m events. Such a strategy must rely only on information from the central detectors. Information from 220m detectors can be included at level one, leading to a higher trigger efficiency for asymmetric 220m + 420m events. 

At ATLAS, the L1 trigger threshold for jets is foreseen to be in the region of 180 GeV and 290 GeV at low and high luminosity respectively. The  reason for the high $E_T$ thresholds is that the QCD jet event rate is large and there is little additional rejection available at level 2 (L2), which can output a maximum rate of $\sim 2$ kHz at ATLAS. It is possible, however, to incorporate information from 420m detectors into L2. The requirement that there are opposite-side forward proton hits at 420m reduces the event rate significantly, as shown in figure \ref{olaplumidep} (a). Furthermore, if timing detectors are used to fix the vertex position, there is an additional rejection of a factor of $\sim20$ as shown in section \ref{fasttiming}. The event rate at the output of L2 can therefore be reduced by requiring two in-time 420m proton hits at L2 by factors of 21500, 570 and 140 for luminosities of $10^{33}$, 5$\times10^{33}$ and $10^{34}$ cm$^{-2}$ s$^{-1}$ respectively. A strategy could be to allow CEP a fixed (pre-scaled if necessary) input rate at L1 for di-jet events with $E_T \geq 40$ GeV. This high L1 rate would then be reduced to an acceptable L2 output rate by proton hit and timing information in the 420m detectors. Further reductions in rate would be possible by cutting on the final state jet variables defined in section \ref{jets}. 

We present results as a function of maximum allowed input L1 jet trigger rate. The maximum L1 input rate we consider is 25kHz, which we label J25. This trigger would be un-prescaled for $40$ GeV di-jets at $10^{33}$ cm$^{-2}$ s$^{-1}$, and be pre-scaled by a factor of $10$ at $10^{34}$ cm$^{-2}$ s$^{-1}$. Given the rejection factors above from requiring two in-time proton hits at L2, the maximum L2 output rate at $10^{34}$ cm$^{-2}$ s$^{-1}$ would be 150Hz. This can be further reduced to a few Hertz at L2 output by applying cuts on the final state jet topology. We also present results for a lower L1 input-rate of 10kHz (J10). 

In addition to the jet trigger, low transverse momentum muon triggers will also save b-jet events. At low luminosity, a single un-prescaled muon trigger with $p_T \geq 6$ GeV is possible. At high luminosity the $p_T$ threshold would be increased to reduce the rate. The rates for the muon triggers could be reduced significantly by requiring a coincidence with a jet with $E_T\geq40$ GeV. We use the notation MU6 for a 6 GeV muon trigger and MU10 for a 10 GeV muon trigger. We assume that no pre-scale is applied, which may mean that the jet requirement is included in the trigger at high luminosity. MU6 will save $11 \%$ of b-jet events. If the threshold were increased to 12 GeV (MU12), $4\%$ of events will be saved.   

At low luminosity, a rapidity gap requirement between the outgoing protons and the central system in addition to two central jets with $E_T \geq 40$ GeV will have an acceptable trigger rate. This trigger is self-prescaling as luminosity increases, because pile-up events will destroy the rapidity gaps. Using a Poisson distribution, we evaluate the probability that there is only one inelastic scatter in the bunch crossing, i.e the event of interest. This rapidity gap trigger provides a luminosity-dependent trigger efficiency of $\sim 17\%$ at $10^{33}$ cm$^{-2}$ s$^{-1}$ and $\sim 2 \%$ at $2 \times 10^{33}$ cm$^{-2}$ s$^{-1}$.

As mentioned above, the 220m detectors can be included in an L1 trigger decision. For instantaneous luminosities up to 2-3$\times10^{33}$~cm$^{-2}$~s$^{-1}$, it is expected that a trigger consisting of one proton hit at 220 m and two 40 GeV central jets requires no pre-scale to have a L1 trigger rate of 1kHz \cite{Grothe:2006dj}. If, as for the di-jet triggers, a relatively large L1 input trigger rate is allowed which can be reduced at L2 by 420m hit information, then it should be possible to achieve a trigger efficiency close to $100\%$ for the asymmetric events at high luminosity, although further studies will be required to confirm this expectation. 

The efficiencies of the above trigger strategies and the effect on the significance of the measurement are presented in section \ref{discussion}.

\subsection{Selection of exclusive di-jet events in the central detector}
\label{jets}
Experimentally, the definition of a central exclusive di-jet event, i.e. an event with no activity outside the jets, is somewhat arbitrary. The CDF collaboration have 
searched for a central exclusive di-jet signal \cite{Terashi:2007kb} using the di-jet mass fraction, $R_{jj}$, where
\begin{equation}
R_{jj} = \frac{M_{jj}}{M},
\end{equation}
where $M_{jj}$ is the mass of the di-jets and $M$ is the mass of the central system given by equation \ref{cepmass} \footnote{CDF do not have proton taggers on each side of the experiment, and therefore estimate $\xi_1$ and $\xi_2$ using the central detector. Details can be found in \cite{Terashi:2007kb}.}. Exclusive events might be expected to have a di-jet mass fraction close to $1$, since there should be no activity outside the jets. However, parton showering, hadronisation and out-of-cone effects (CDF use a cone-based jet algorithm) smear the measurement, so that even for true 
exclusive di-jet events, $R_{jj}$ can be significantly less than $1$ \cite{Cox:2005gr}.

An alternative definition of the di-jet mass fraction was proposed in \cite{Khoze:2006iw} which aims to reduce the effects of hard final state radiation. In a central exclusive event, the highest transverse energy jet is the one that has been least affected by final state radiation. The di-jet mass fraction $R_j$ is defined as  
\begin{equation}
R_j = \frac{2E_{T}^{1}}{M} \textrm{cosh} (\eta_1 - y),
\end{equation}
where $E_{T}^{1}$ and $\eta_1$ are the transverse energy and pseudo-rapidity of the leading jet in the event. 
The pseudo-rapidity of the central system, $y$, is obtained from the measurement of the forward protons;
\begin{equation}
y= \frac{1}{2} \textrm{ln} \left( \frac{\xi_1}{\xi_2} \right).
\end{equation}
The pseudo-rapidity of the central system can also be obtained from the average pseudo-rapidity of the two jets. This means that the variable
\begin{equation}
\Delta y = y - \left( \frac{\eta_1 + \eta_2}{2} \right)
\end{equation}
should be approximately zero for a central exclusive event.

\subsection{Jet algorithms}
\label{sec:jets}

In this section we investigate the dependence of the jet-based variables $R_{jj}$, $R_j$ and $\Delta y$ on the choice of jet algorithm. We use the mid-point cone algorithm \cite{Blazey:2000qt} with varying cone radius and the $K_T$ algorithm \cite{Butterworth:2002xg} with different values of the R-parameter. Events are accepted if the highest transverse energy jet has $E_T \geq 40$ GeV.

\subsubsection{Generator level}
\begin{figure}
\centering
\mbox{
	\subfigure[]{\includegraphics[width=.5\textwidth]{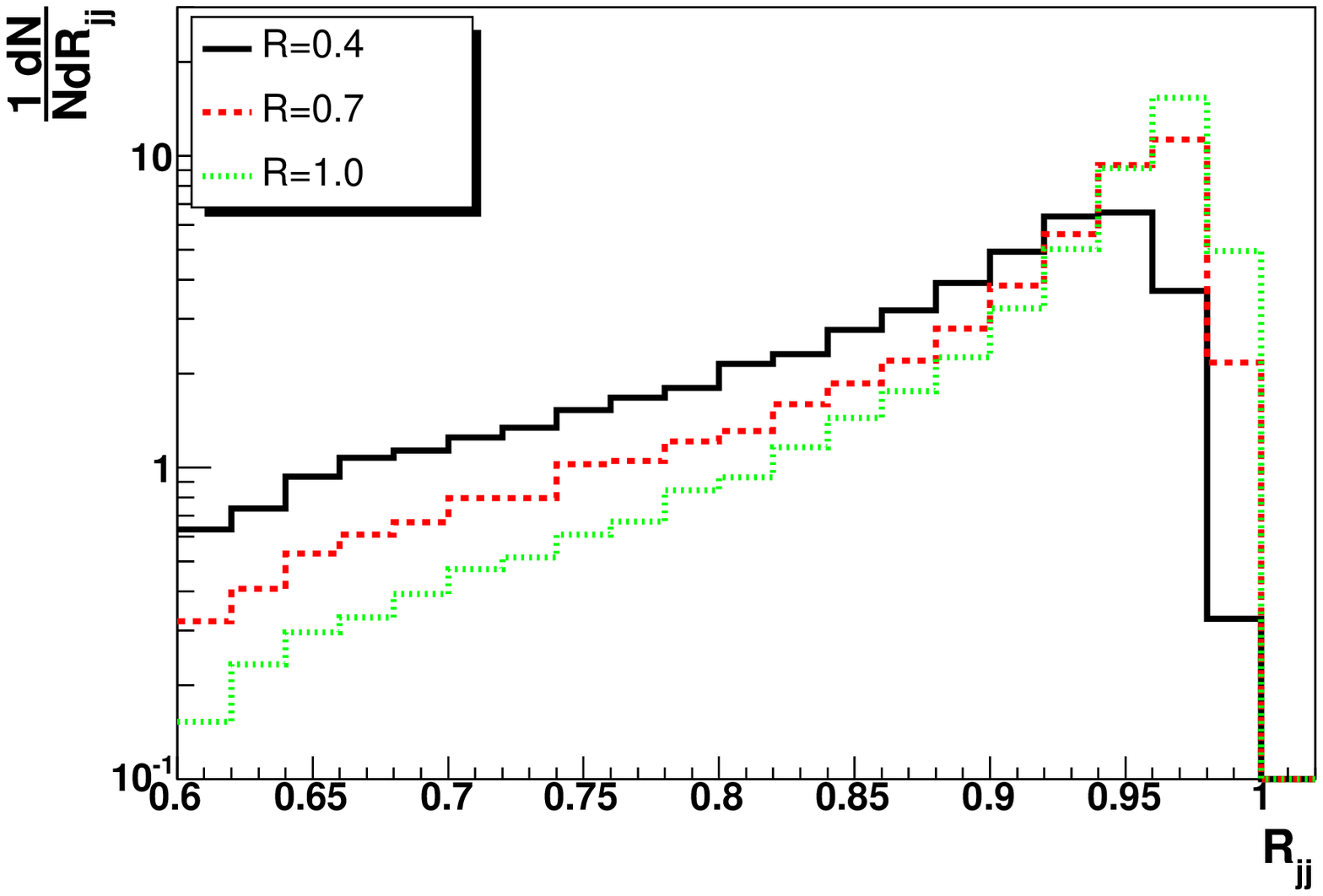}} \quad
	\subfigure[]{\includegraphics[width=.5\textwidth]{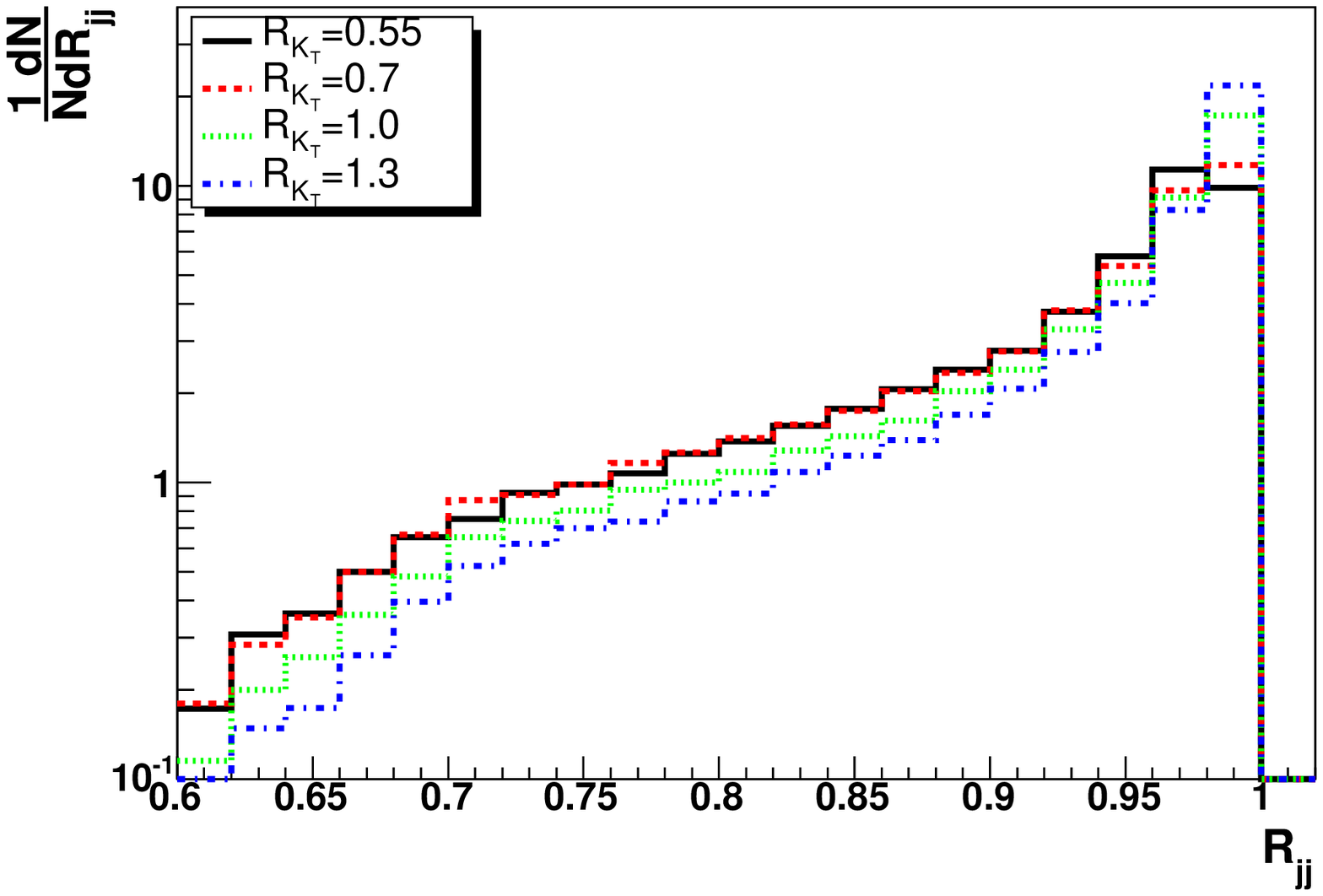}}
}
\mbox{
	\subfigure[]{\includegraphics[width=.5\textwidth]{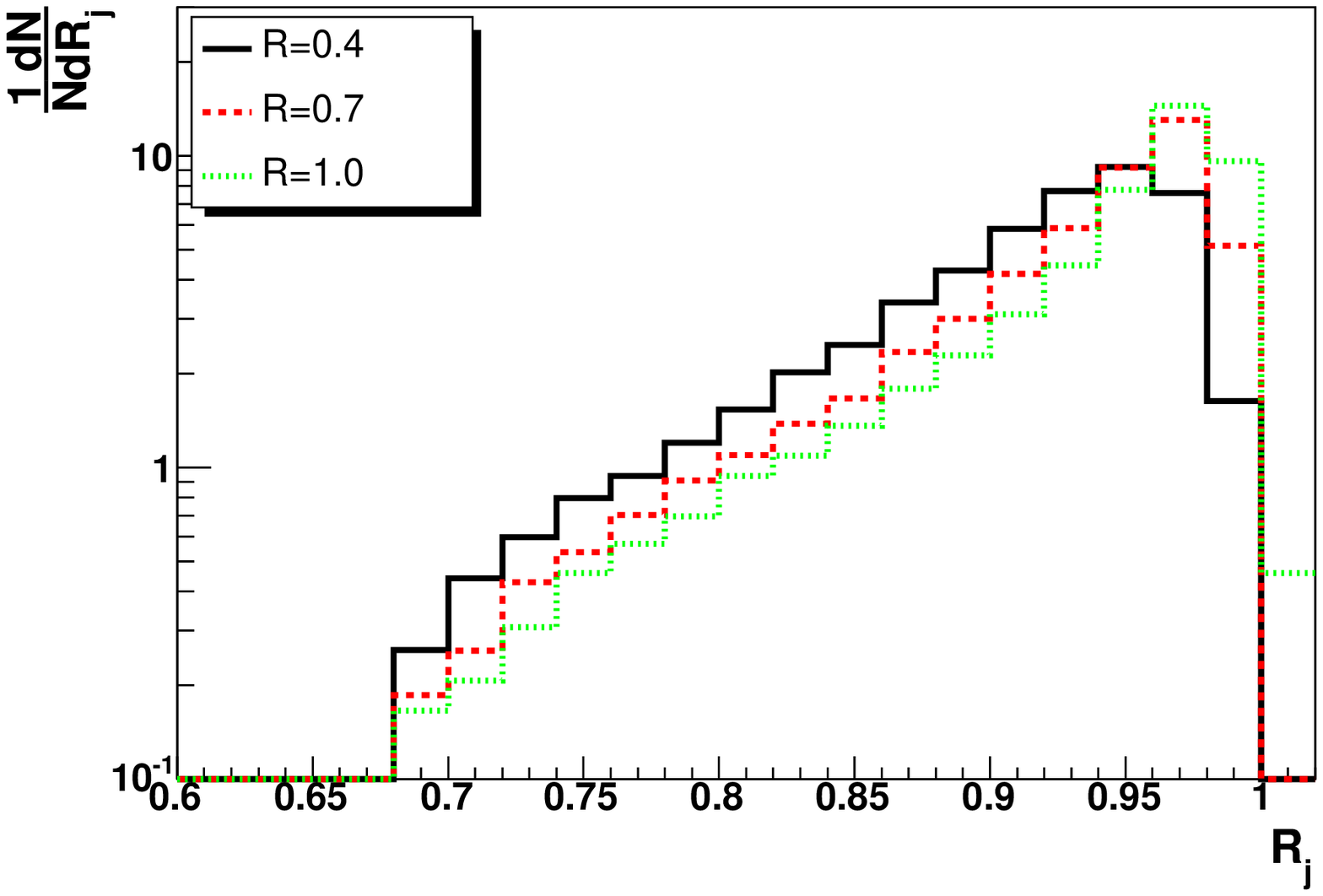}} \quad
	\subfigure[]{\includegraphics[width=.5\textwidth]{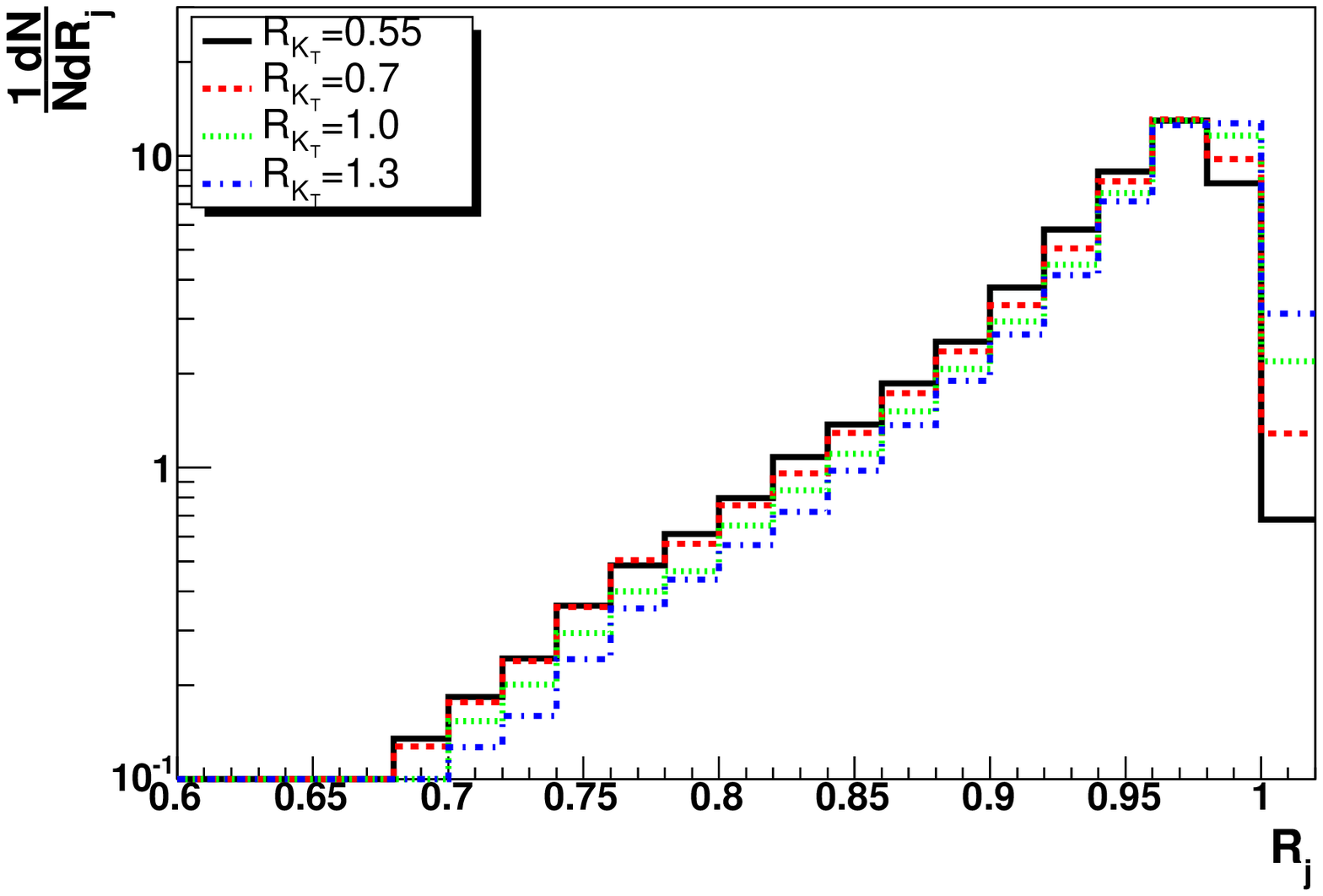}}
}
\mbox{
	\subfigure[]{\includegraphics[width=.5\textwidth]{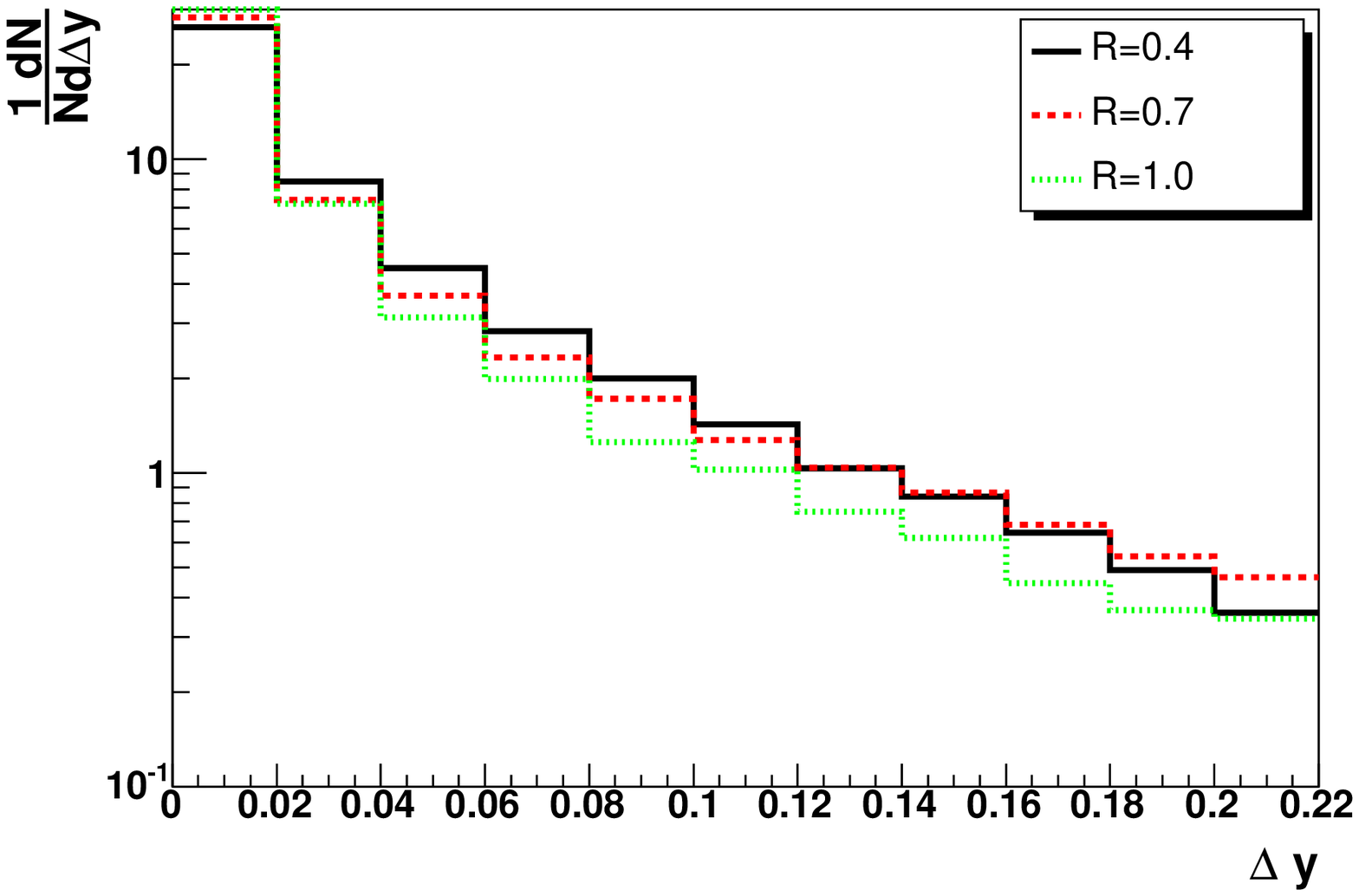}} \quad
	\subfigure[]{\includegraphics[width=.5\textwidth]{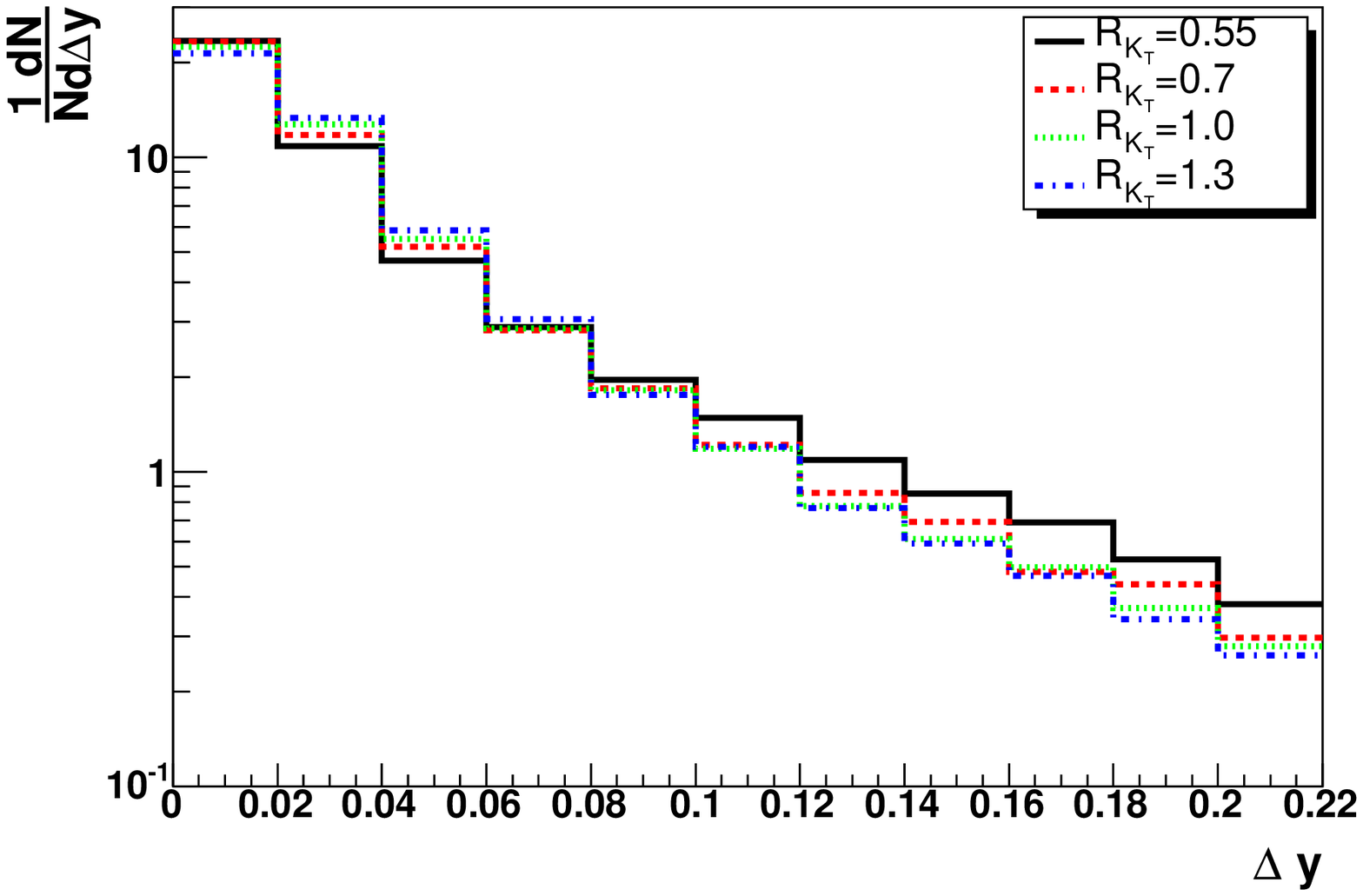}}
}
\caption{The dependence of the $R_{jj}$, $R_j$ and $\Delta y$ variables on the jet algorithm used to reconstruct the jets. Figures (a), (c) and (e) show the dependence of these variables on the radius used in the cone algorithm. Figures (b), (d) and (f) show the dependence on the $K_T$ di-jet. \label{SignalPlotsHadron}}
\end{figure}

The $R_{jj}$ distributions are shown in figures \ref{SignalPlotsHadron} (a) and \ref{SignalPlotsHadron} (b) for the cone and $K_T$ algorithms respectively. As expected, the distributions peak at $R_{jj} \sim 1.0$ with a tail that extends to lower values due to parton showering and hadronisation effects. As the cone radius, $R$, is decreased, the distribution is smeared to lower values of $R_{jj}$ because for smaller cone radii, particles are more likely to be reconstructed outside of the jets and the reconstructed mass of the di-jet system is correspondingly lower. A similar result is found for the $K_T$ algorithm when reducing the R-parameter, $R_{K_T}$. 

The dependence on the cone radius and R-parameter is less marked for the $R_j$ variable. The distributions are still peaked at $R_j \sim 1.0$, but changing the cone radius or R-parameter has only a small effect on the distribution as shown in figures \ref{SignalPlotsHadron} (c) and \ref{SignalPlotsHadron} (d). This is expected because the leading jet is less affected by final state radiation, and should therefore be less affected by changes in jet definition.

The $\Delta y$ distributions are shown in figures \ref{SignalPlotsHadron} (e) and \ref{SignalPlotsHadron} (f) for the cone and $K_T$ algorithms respectively. The distributions are strongly peaked at zero for both jet algorithms and all the choices of cone radius and R-parameter result in the majority of events being reconstructed in the region $\Delta  y \leq 0.06$. We conclude that the $\Delta y$ and $R_j$ variables are relatively insensitive to the choice of jet algorithm before detector effects are taken into account.

\subsubsection{Detector effects}
\label{detector}

\begin{figure}
\mbox{
	\subfigure[]{\includegraphics[width=.5\textwidth]{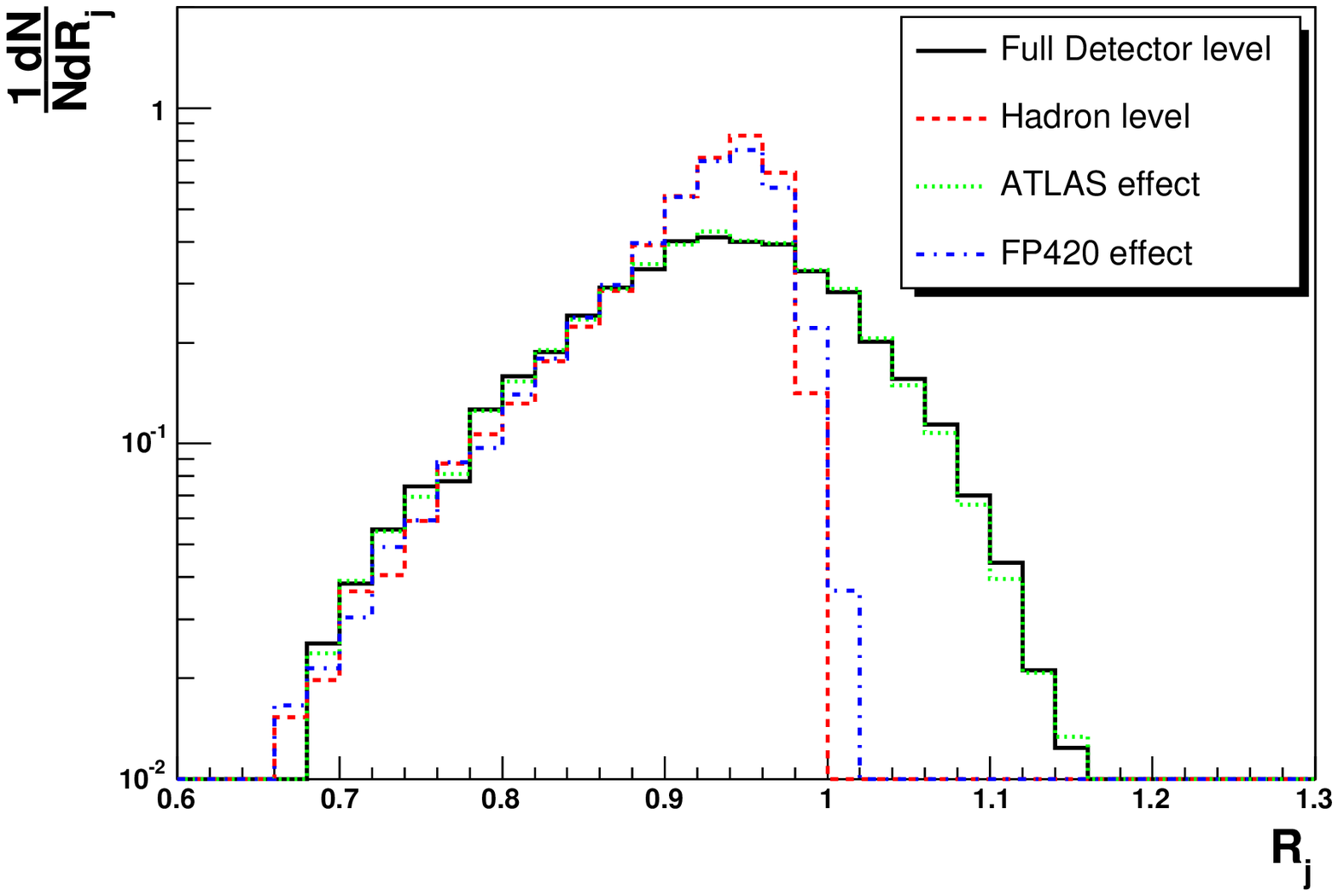}} \quad
	\subfigure[]{\includegraphics[width=.5\textwidth]{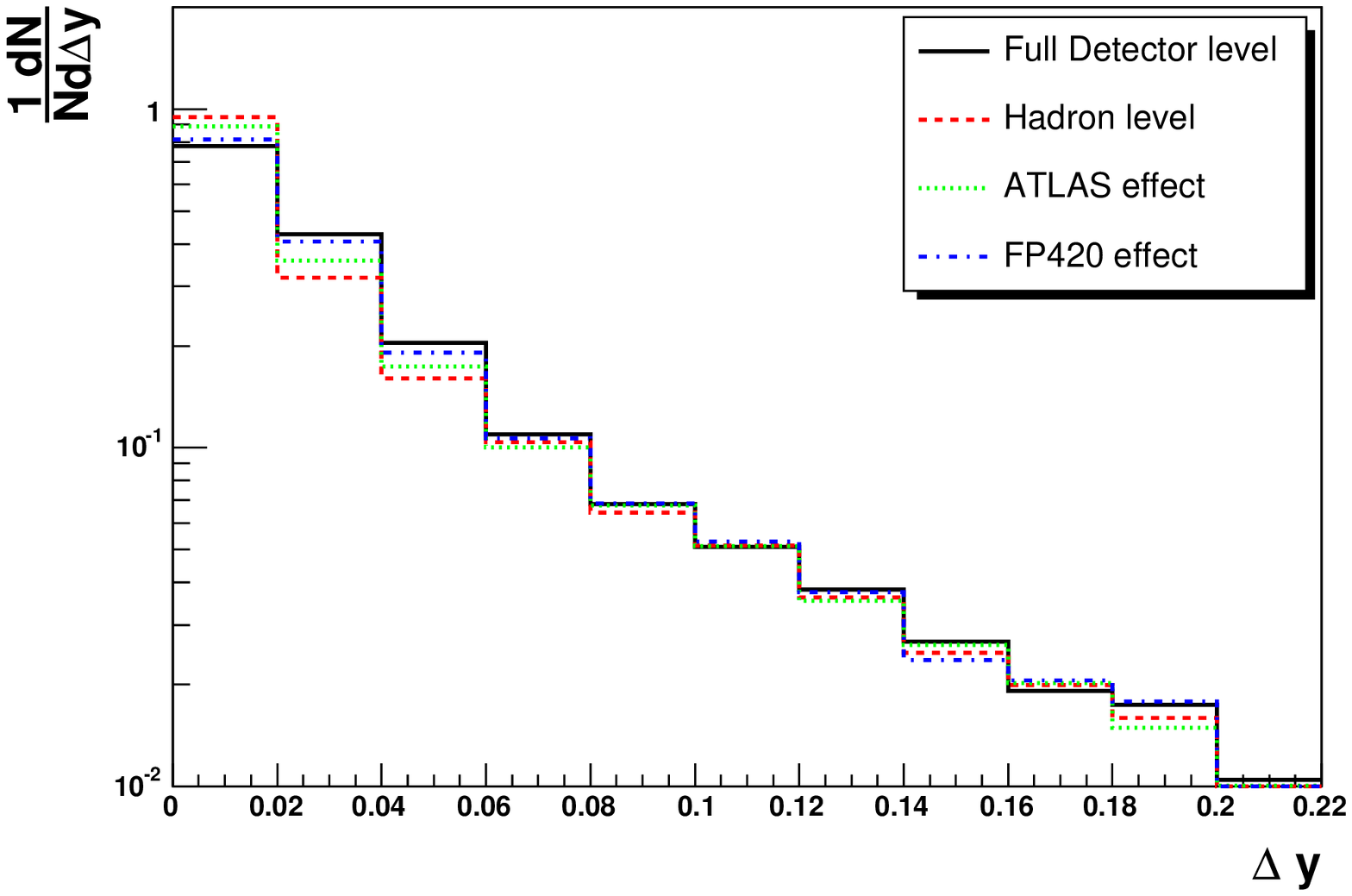}}
}
\caption{The effect of detector resolution on the $R_j$ and $\Delta y$ variables of the reconstructed signal events. The resolution of the ATLAS calorimeters dominates the smearing of the $R_j$ variable (figure (a)) and the resolution of FP420 only has a small effect. Figure (b) shows that the $\Delta y$ variable is only slightly affected by the detector resolution. \label{signaldeteffects}}
\end{figure}

Figure \ref{signaldeteffects} (a) and (b) show the effects of detector smearing on the $R_j$ and $\Delta y$ distributions for CEP Higgs events (i.e. signal only) for a cone radius of 0.4. The particle energy, pseudo-rapidity and azimuth are smeared by the resolution of the ATLAS electromagnetic and hadronic calorimeters \cite{calorimeters}. The momentum of the leading protons measured by the forward detectors is smeared by the resolution of the 420m detectors \cite{Bussey:2006vx}. 

The $R_j$ distribution is smeared around the peak value. The effect of the ATLAS central detector resolution on the $E_T$ measurement of the leading jet is the dominant effect, with the smearing of the mass measurement from the 420m detectors contributing only a small amount. For the $\Delta y $ distribution, the detector effects are minimal.

In addition to smearing, the $b$-tagging efficiency of the ATLAS detector must be taken into account. The single $b$-tagging efficiency at ATLAS is 60\% for a $b$-jet, with a mis-tag rate of 1.3\% for jets originating from gluons \cite{btag}. Therefore the efficiency for a system consisting of two $b$-jets being  experimentally identified as a two $b$-jet system is 0.36. The mis-tag rate for a gluon and $b$ di-jet is 0.0078 and the mis-tag rate for a two gluon di-jet is 1.69$\times$10$^{-4}$. The c-quark mis-tag rate is 0.1.

\subsection{Identification of diffractive events}\label{iddiff}

Diffractive events have historically been identified by searching for a rapidity gap between the central system and the outgoing proton. This approach has a high efficiency at the LHC only at very low luminosities, (around  $10^{32}$~cm$^{-2}$~s$^{-1}$). At higher luminosities, particles from pile-up events will in most cases destroy the rapidity gap. As stated in section \ref{trigger}, the probability for no inelastic pile-up events is $\sim$17\% at an instantaneous luminosity of $10^{33}$~cm$^{-2}$~s$^{-1}$, reducing to $\sim$2\% at $2\times10^{33}$~cm$^{-2}$~s$^{-1}$. 

Because of this, we use charged track multiplicity cuts to identify CEP candidate events. The primary hard scatter in a non-diffractive overlap event is characterised by a central system that is colour connected to the proton remnants. There will therefore be on average more charged particles associated with the di-jet vertex in an inelastic collision than in a CEP collision. Furthermore, the underlying event, which cannot be present in a CEP event, will be an additional source of charged particles at the vertex. At ATLAS, the inner detector has excellent charged track vertex reconstruction, which allows charged tracks to be identified with a specific di-jet vertex. We use two variables to identify CEP candidate events using the charged particles that are associated with the di-jet vertex: 
\begin{itemize}
\item{The number of charged particles, $N_C$, that are not contained within the two highest $E_T$ jets.} 
$N_C$ is a measure of the particle multiplicity in the event that is not associated with the hard scatter. It is of course dependent on the jet algorithm used to reconstruct the jets.
\item{The number of charged particles, $N_C^{\perp}$, that are perpendicular in azimuth to the leading jet.}
$N_C^{\perp}$ is a measure of the particle multiplicity associated with the underlying event. There is no underlying event in the CEP signal events.  
We use the definition adopted in \cite{Field:2006gq}, which assigns charged particles to the underlying event if they satisfy
\begin{equation}\label{transverse}
\frac{\pi}{3} \leq | \phi_k - \phi_j | \leq \frac{2\pi}{3} \quad \textrm{and} \quad 
\frac{4\pi}{3} \leq | \phi_k - \phi_j | \leq \frac{5\pi}{3}
\end{equation}
where $\phi_k$ is the azimuthal angle of the charged particle and $\phi_j$ is the azimuthal angle of the highest transverse energy jet.
\end{itemize}

\subsection{Experimental cuts on the final state}

\begin{figure}
\centering
\mbox{
	\subfigure[]{\includegraphics[width=.5\textwidth]{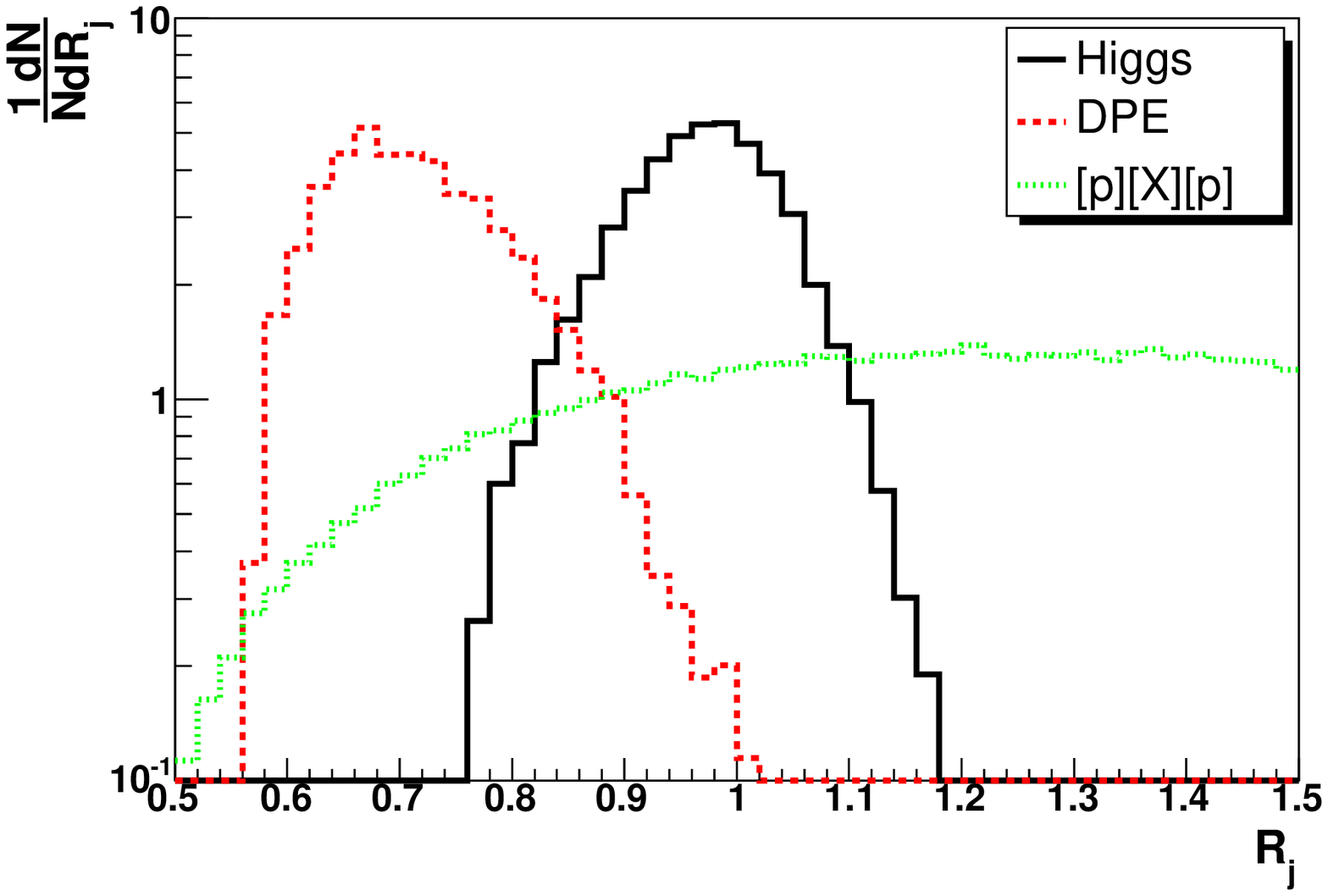}} \quad
	\subfigure[]{\includegraphics[width=.5\textwidth]{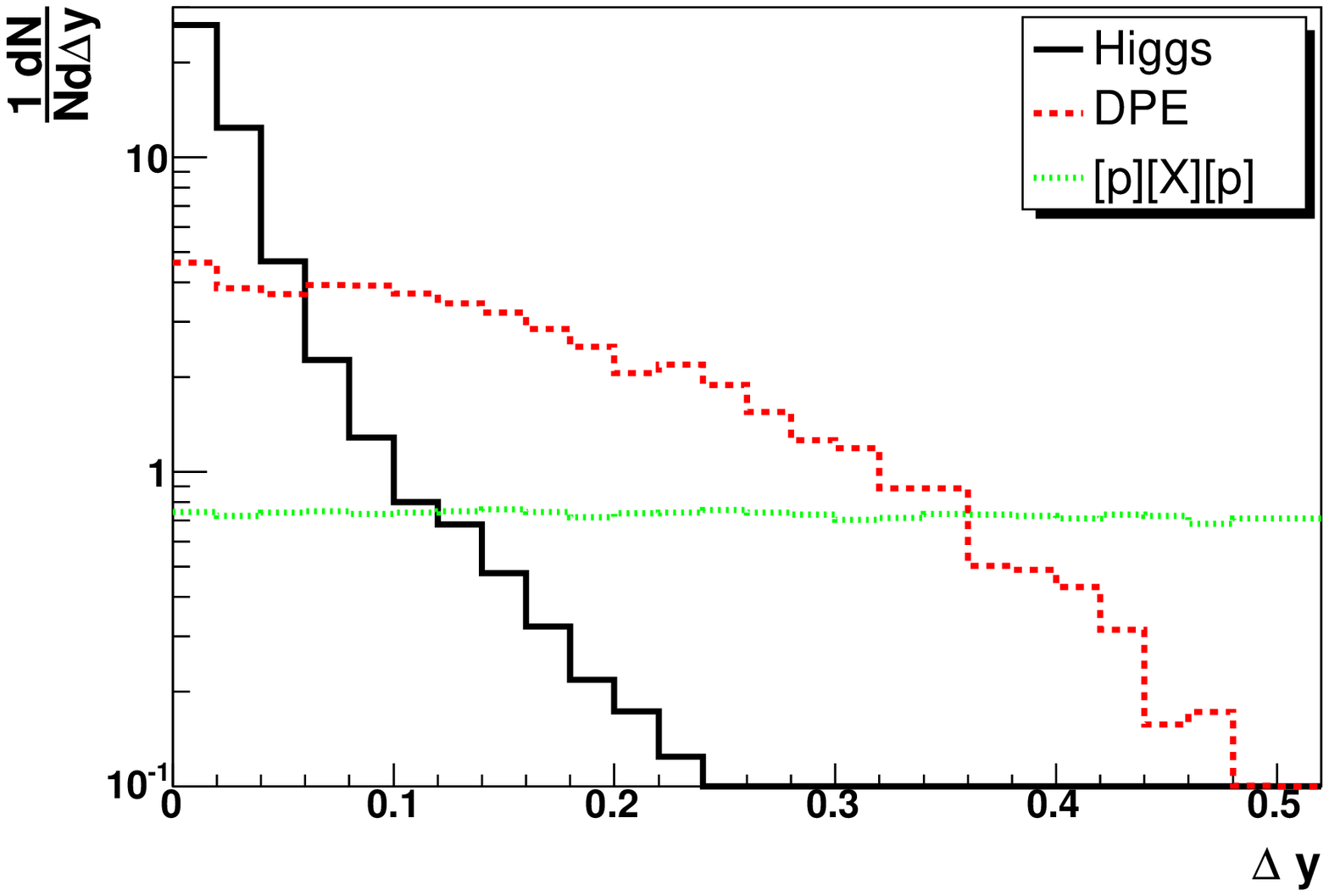}}
	}
\caption{The $R_j$ and $\Delta y$ distributions are shown in (a) and (b) respectively for the signal, [p][$b\bar{b}$][p] and DPE [p $b\bar{b}$ p] backgrounds . The distributions were reconstructed using a cone radius of 0.7 after smearing the particles with detector resolution. \label{signalvbackgroundrjdely}}
\end{figure}

Having defined the experimental variables, we now turn to developing a set of cuts that will maximise the significance of the CEP signal. We begin with the variables, $R_j$ and $\Delta y$. In figure \ref{signalvbackgroundrjdely}, the distributions for signal events are compared with the $b\bar{b}$ backgrounds from double pomeron exchange and overlap ([p][X][p]) events. A cone radius of 0.7 is used and the distributions have been normalised to unity. The overlap background extends to large values of $R_j$ because there is no correlation between the mass as measured in the 420m detectors and the central di-jet mass. The DPE background events have a lower average $R_j$ because of the necessary presence of pomeron remnants which will often lie 
outside of the jet cones. We define an exclusive-enriched sample by $0.75 \leq R_j \leq 1.1$,  $\Delta y \leq 0.06$. The upper limit of the $R_j$ window is chosen to reject overlap background, and should therefore be optimised for instantaneous luminosity.

We now focus on the rejection of inclusive overlap background, [p][X][p], utilising the charged track multiplicity variables described in section \ref{iddiff}. All the distributions shown have the cuts on $R_j$ and $\Delta y$ applied as described above. Charged particles are required to satisfy $p_T \geq0.5$ GeV and $|\eta| \leq 1.75$. To take into account the track reconstruction efficiency of the ATLAS inner detector for very low $p_T$ tracks, 20\% of the generated charged tracks outside of the di-jets are disregarded at random. Experimentally, charged tracks would only be used if they are associated with the di-jet vertex. The approach we adopt is to only consider tracks that fall within a vertex window of width $\pm \delta$ mm around the di-jet vertex. $\delta$ must be larger than the vertex resolution of the inner detector and we choose
\begin{equation}
\delta = \pm 2 \, \sigma_z
\end{equation}
where $\sigma_z$ is the vertex resolution of the most poorly reconstructed particle. The vertex resolution of the inner detector may be parametrised \cite{innerdetector} by 
\begin{equation}
\label{vres}
\sigma_z  = 87 + \frac{115}{p_T\sqrt{\sin^3\theta}} \quad (\mu\rm{m}).
\end{equation}
Particles with low transverse momentum or large pseudo-rapidity have poorer vertex resolution and hence a larger vertex window is applied. 

\begin{figure}
\centering
	\includegraphics[width=.5\textwidth]{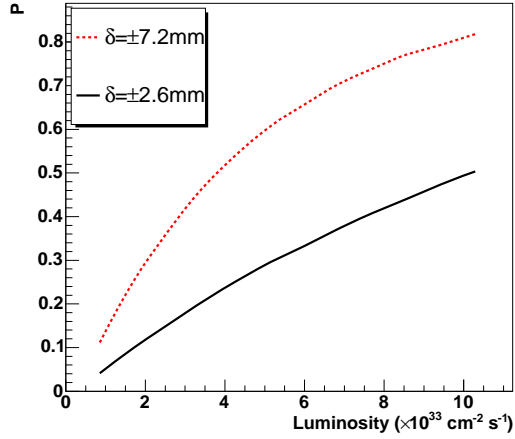}
\caption{The probability of at least one inelastic pile-up event falling within a given distance, $\delta$, of the primary event vertex. \label{inelasticprob}}
\end{figure}

Figure \ref{inelasticprob} shows the probability of an inelastic event falling within $\pm \delta$ mm of the primary vertex. From equation \ref{vres}, a particle with $p_T=0.5$ GeV and $\eta = 2.5$ requires a vertex window of $\delta = \pm 7.2$ mm and therefore, at $10^{34}$~cm$^{-2}$~s$^{-1}$, 80\% of events will have at least one inelastic pile-up event inside the vertex window. Any cuts on the charged particle multiplicity variables will therefore be relatively ineffective. This is the reason why the pseudo-rapidity of particles used in the charged particle multiplicity cuts is restricted to $|\eta| \leq 1.75$, which allows for a smaller vertex window of $\pm2.6$ mm for a particle with $p_T=0.5$ GeV. 



\begin{figure}
\centering
\mbox{
	\subfigure[]{\includegraphics[width=.5\textwidth]{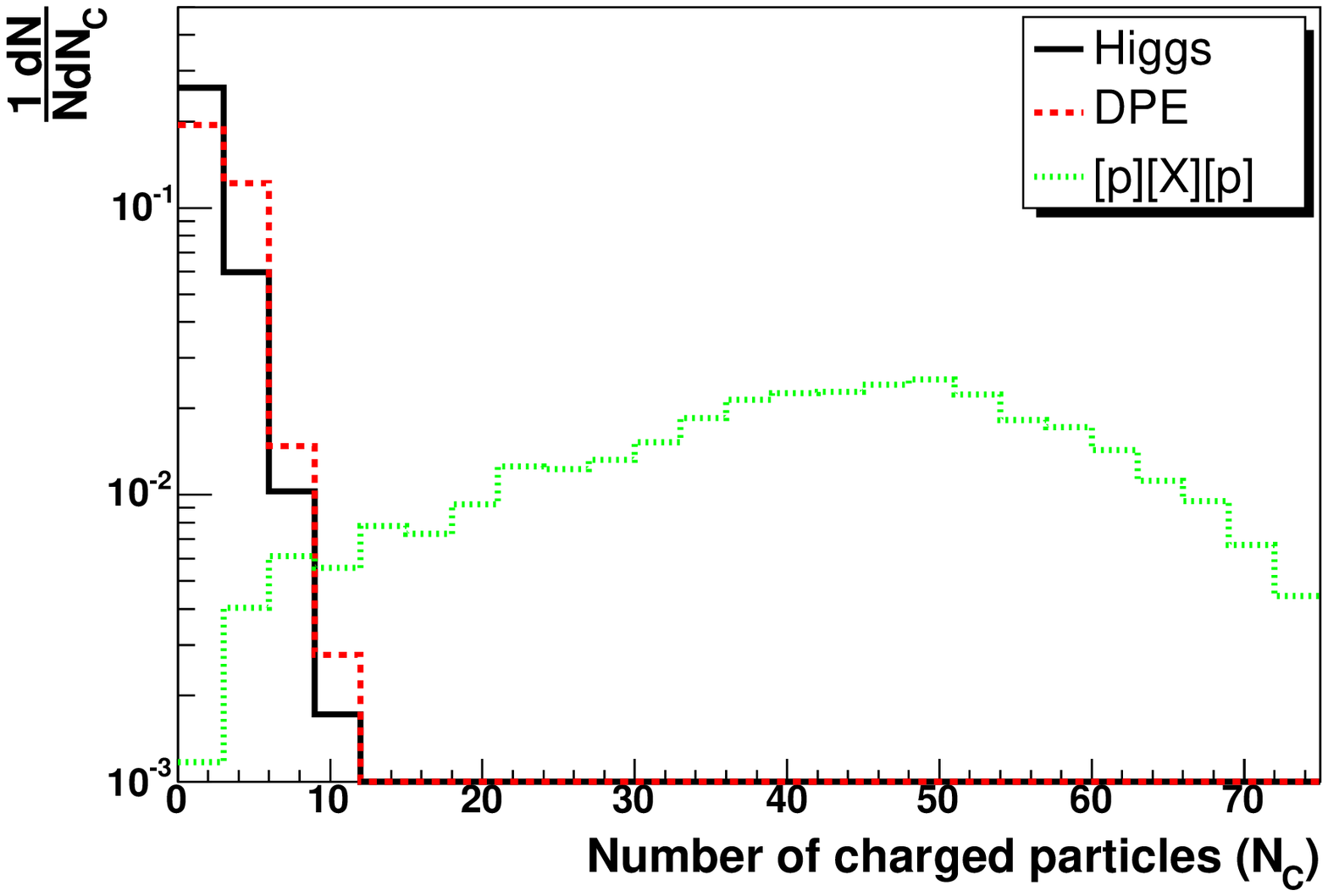}} \quad
	\subfigure[]{\includegraphics[width=.5\textwidth]{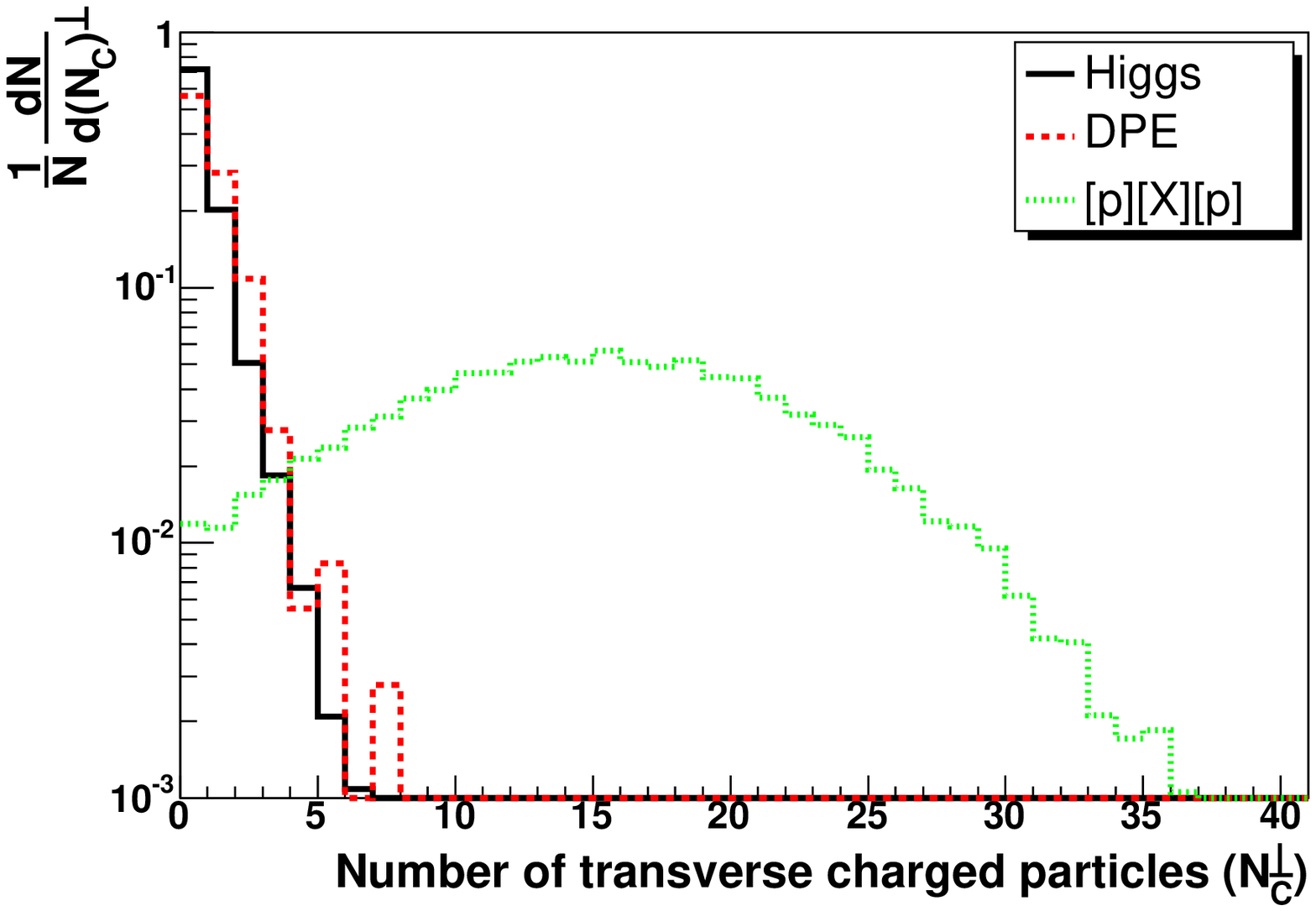}}
	}
\caption{The charged track multiplicity outside of the di-jet system, $N_C$, is shown in figure (a). Figure (b) shows the number of charged particles that are transverse to the leading jet as defined by equation \ref{transverse}. In both cases the particles must satisfy $p_T>0.5$ GeV and $|\eta\leq1.75$. Only 80\% of the particles are used to replicate ATLAS reconstruction efficiency for low $p_T$ tracks. \label{ncharged}}
\end{figure}


Figure \ref{ncharged} (a) shows the number of charged particles outside of the di-jet system, but associated with the di-jet vertex, for the signal, DPE $b\bar{b}$ and [p][X][p] $b\bar{b}$ events reconstructed using the cone algorithm with a radius of 0.7. The diffractive events (CEP and DPE) have few tracks outside of the di-jet system, whereas the inclusive overlap background has many more charged particles.
Figure \ref{ncharged} (b) shows the transverse charged particle distributions, $N_C^{\perp}$. The exclusive events are more strongly peaked at zero in this distribution. This is as expected because $N_C^{\perp}$ is designed to measure the underlying event activity. Similar results are observed for both charged particle distributions using the $K_T$ algorithm with $R_{K_{T}}=0.55$.

\begin{figure}
\centering
	\includegraphics[width=.5\textwidth]{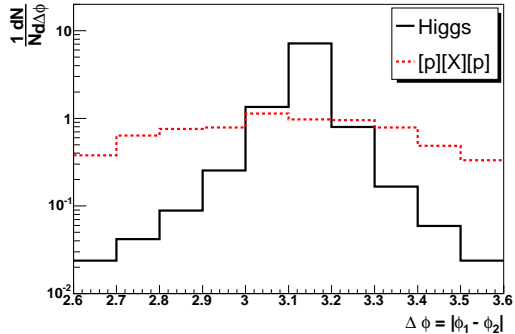}
\caption{The azimuthal angle between the two leading jets for the signal and [p][$b\bar{b}$][p] background. \label{deltaphi}}
\end{figure}

The final central detector variable that we use to reject inclusive overlap events is $\Delta \phi$, the angle between the two leading jets in the event. Central exclusive events have no initial state radiation (ISR) and so the di-jet system is produced with no transverse boost. This means that the jets will be back-to-back in azimuth, i.e $\Delta \phi \sim \pi$. Inclusive overlap events can have a large transverse boost due to ISR. Figure \ref{deltaphi} shows the $\Delta \phi$ distribution for the signal and [p][X][p] $b\bar{b}$ events. The distribution is again made after the $R_j$ and $\Delta y$ cuts which remove the majority of events that are not back to back due to wide angle, final state radiation. We apply the cut $\pi - |\Delta \phi|  \leq 0.15$ to further enrich the exclusive sample.  

\subsubsection{Overlap background rejection using fast timing detectors}\label{fasttiming}

Each of the overlap backgrounds can be reduced significantly by matching the di-jet vertex to the event vertex reconstructed from the proton time-of-flight information. The proposed 420m and 220m detectors at the LHC each have fast timing components for this purpose. As we shall see, the timing resolution is critical for rejecting overlap background at the highest LHC luminosities. The aim is to achieve a time-of-flight measurement with a resolution of 10ps \cite{Albrow:2005ig}. We take this figure as the benchmark in what follows.     

The vertex position for overlap events reconstructed by the forward detectors may not coincide in space or time with the di-jet vertex reconstructed in the central detector. To calculate the timing rejection factor for [p][X][p] events we select three `event vertices' distributed in longitudinal position according to a gaussian distribution of width 4.45cm and time according to a gaussian distribution of width 0.18ns \cite{White:2007ha}. We define the first vertex to be the di-jet vertex, and the second and third to be the soft-single diffractive vertices that produce the outgoing protons. The fake vertex position $z_f$ as measured by the forward detectors is given by 
\begin{equation}\label{tofvertex}
z_f = \frac{c}{2}\left(t_{-} - t_{+} \right) 
\end{equation}
where the $t_{\pm}$ are the arrival times of the protons measured in the forward detectors in the $\pm$ z direction. For overlap events made up of two proton-proton interactions (i.e. [p][pX] and [pp][X]) a similar method can be used after selecting two event vertices (the first of which is the primary di-jet vertex).  An event is kept if the fake vertex lies within $\pm4.2$ mm of the primary vertex (this is $\pm 2\sigma$ assuming a timing resolution in the forward detectors of 10ps).

\begin{figure}
\centering
\includegraphics[width=.5\textwidth]{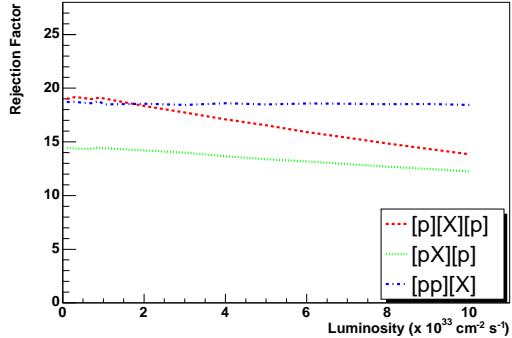}
\caption{Time-of-flight rejection factor for each type of overlap background, assuming a timing resolution of 10ps in the forward detectors.\label{tofrejection}}
\end{figure}

We must also take account of the possibility that more than one proton might be detected in each forward detector. In this case, if any combination of protons produces a fake vertex within the $\pm4.2$ mm window then they will survive the vertex cuts. This implies that the rejection factor decreases at higher luminosities where the probability for additional protons is larger. Figure \ref{tofrejection} shows the rejection factor for each type of background as a function of luminosity. 

The rejection factors presented in figure \ref{tofrejection} correspond to a 10ps timing resolution, which is the base design of FP420. It should be noted that if this timing resolution is improved, the rejection factors increase linearly with reduction in time-of-flight measurement, i.e the rejection factors increase by a factor of 5 for 2ps timing resolution. Furthermore, it has been proposed that the central detector could provide a time of interaction measurement for the primary vertex \cite{White:2007ha}. In such a case, the time of arrival of each proton can be compared to the time of the interaction and this results in a order of magnitude increase in rejection for the [p][X][p] overlap background if the central interaction is measured to a 10ps accuracy.

\section{Results}
\label{sec:results}
\subsection{Standard Model Higgs boson}\label{smresults}

We focus initially on the symmetric CEP events, i.e. those for which both protons are detected in 420m proton detectors at each side of the central detector. CEP signal and di-jet background events are generated by ExHuME, and added to the non-CEP di-jet backgrounds generated by HERWIG + JIMMY and POMWIG, with forward protons added into the event record in the case of overlap events as described in section \ref{backgrounds}. All Monte Carlo events will therefore have two forward protons, one on each side of the central di-jet system. The 4-momenta of all outgoing particles are smeared to simulate the response of the ATLAS detector, as described in section \ref{detector}. Jets are found using the midpoint cone algorithm with a cone radius of 0.7 (section \ref{sec:jets}). The top line of table \ref{cuteff} shows the cross sections for each sample, after the following cuts on the Monte Carlo events:
\begin{itemize}
\item
Both forward protons must lie within the geometric acceptance of the 420m forward detectors, as described in section \ref{fp420}, with the detectors positioned 5mm from the beam. This corresponds approximately to the kinematic range $0.005 \leq \xi_1 \leq 0.018$, $0.004 \leq \xi_2 \leq 0.014$ and unrestricted in~$t$. 
\item The mass of the central system as measured in the forward detectors must satisfy
$80 {\rm~GeV} \leq M \leq 160$ GeV.
\item 
Two jets with $E_{T1} \geq 45$ GeV, $E_{T2} \ge30$ GeV.
\end{itemize}   
The exclusive candidate sample is defined by the following cuts:
\begin{itemize}
\item The di-jet mass fraction $0.75 \leq R_j \leq 1.1$. 
\item The rapidity of the central system $\Delta y \leq 0.06$. 
\item The jets must be back-to-back in azimuth, i.e $\pi - |\Delta \phi|  \leq 0.15$.
\item The multiplicity of charged tracks with $p_T \geq 0.5$ GeV and $|\eta| \leq 1.75$ that are associated with the di-jet vertex, $N_C \leq 3$ and $N_C^{\perp} \leq 1$. In order to simulate the ATLAS detector reconstruction efficiency, only 80\% of charged particles are used. 
\end{itemize}

\begin{table}[t]
\centering
\begin{tabular}{|c||c|c|c||c||c|c|c|}
\hline
& \multicolumn{7}{|c|}{Cross section (fb)} \\
\hline
Cut & \multicolumn{3}{|c||}{CEP} & DPE & [p][X][p] & [p][pX] & [pp][X] \\
\hline
& $H$ & $b\bar{b}$ & $gg$ & $b\bar{b}$ & $b\bar{b}$  & $b\bar{b}$ & $b\bar{b}$ \\
\hline
 $E_T$, $\xi_1$, $\xi_2$ & 0.124 & 1.320 & 2.038 & 0.633 & 3.91$\times10^{5}$& 7.33$\times10^{2}$ & 6.29$\times10^{4}$  \\
 $R_j$ &  0.119 & 1.182 & 1.905 & 0.218 & 4.73$\times10^{4}$ & 85.2 & 7.59$\times10^{3}$ \\
 $\Delta y$ & 0.010 & 1.036 & 1.397 & 0.063 & 2.16$\times10^{3}$ & 1.38 & 3.50$\times10^{2}$ \\
 $\Delta \Phi$ & 0.093 & 0.996 & 1.229 & 0.058 & 6.66$\times10^{2}$ & 0.77 & 1.07$\times10^{2}$\\
 $N_C$, $N_C^{\perp}$ & 0.084 & 0.923 & 0.932 & 0.044 & 6.49 & 0.45 & 1.35 \\
 $\Delta$M & 0.072 & 0.070 & 0.084 & 0.004 & 0.59 & 0.03 & 0.13 \\
\hline
\end{tabular}
\caption{Cross section (fb) after applying each cut for the analysis performed using the cone algorithm with radius 0.7. The first cut is applied after requiring that both protons are tagged at 420m, the mass measured by the forward detectors is between 80 and 160 GeV and the transverse energy of the leading jet is greater than 40 GeV. The overlap backgrounds are defined at high luminosity (10$^{34}$~cm$^{-2}$~s$^{-1}$) and TOF vertexing has been used. \label{cuteff}}
\end{table}%

Table \ref{cuteff} shows the effect of the cuts on the signal and background samples. For the overlap and DPE backgrounds, only the dominant $b \bar b$ contributions are shown. The contribution from the non $b$-jet backgrounds are shown after all analysis cuts in table \ref{crosssectionfinal}.
The final cut in table 2 is a mass window, $\Delta M$, around the Higgs boson peak;
\begin{equation}\label{masswin}
\Delta M = M \pm \sqrt{2}\left(\sigma_M^2 + \Gamma_H^2 \right)^{\frac{1}{2}}
\end{equation}
where $\Gamma_H$ is the width of the Higgs boson and $\sigma_M$ is the mass resolution of the forward detectors. A 120 GeV SM Higgs boson has a very small width, and the size of the mass window is therefore dominated by the resolution of the forward detectors. The size of the mass window is chosen so that more than 80\% of the signal is retained. The mass window for events in which both protons are tagged at 420m is $\pm$2.4~GeV. 

The largest loss of signal events is caused by the double $b$-tagging efficiency of 0.36 and the acceptance of the forward detectors. It may be possible to improve the $b$-tagging efficiency for central exclusive events due to the clean nature of the CEP vertex. In addition, various strategies have been discussed to increase the acceptance of the proposed 420m detectors \cite{Albrow:2005ig}.

The final cross sections for both signal and backgrounds are shown in table \ref{crosssectionfinal}. Note that we do not include the trigger efficiency in this table. The final results for the two jet algorithms are similar and we use only the cone algorithm for the remainder of this paper. The mass window defined by equation \ref{masswin} has been applied so that the size of the signal and background in the region of interest can be evaluated. At low luminosity, the dominant backgrounds are from central exclusive di-jet production. At high luminosity the [p][X][p] background becomes dominant. In table \ref{crosssectionfinal} we also show the signal and background cross sections for the asymmetric events for a distance of approach of (5mm + 2mm) and (3mm + 1.5mm) of the 420m and 220m detectors from the beams. Because of the poorer resolution of the 220m taggers, the mass window is increased to $\pm$4.9~GeV\footnote{The achievable mass resolution for asymmetric events is slightly different on each side of the IP due to a difference in beam 1 and beam 2 optics. Here we quote the average mass resolution.}.
    
\begin{table}[t]
\centering
\begin{tabular}{|c|c||c|c||c|c|}
\hline
Generator & Process & \multicolumn{2}{c|}{$\sigma_{420-420}$ (fb)} & \multicolumn{2}{c|}{$\sigma_{420-220}$ (fb), CONE} \\
&  & $K_T$     & CONE & 5mm/2mm & 3mm/1.5mm \\
\hline
\hline
ExHuME & $H\rightarrow b\bar{b}$ & 0.071 & 0.072 & 0.038 & 0.115 \\ %
\hline
\hline
ExHuME & $b\bar{b}$ & 0.076 & 0.070 & 0.067 & 0.203 \\
 & $gg$ & 0.066 & 0.084 & 0.091 & 0.278 \\ %
 \hline
 \hline
POMWIG  & $b\bar{b}$&  0.011 & 0.004  & 0.004 & 0.013 \\ %
&  $jj$  & 0.0005 & 0.0002 & 0.0002 & 0.0007\\ %
\hline
\hline
$\rm{[p][X][p]}$ (L) & $b\bar{b}$ & 0.0029 & 0.0037 & 0.0032 & 0.0097\\ 
 & $jj$ & 0.0003 & 0.0003 & 0.0003 & 0.0009\\ 
$\rm{[p][X][p]}$ (H)  & $b\bar{b}$  & 0.46 & 0.59 & 0.46 & 1.41 \\ 
 &  $jj$  & 0.04 & 0.05 & 0.04 & 0.13 \\     
\hline
$\rm{[pp][X]}$ (L) & $b\bar{b}$ & 0.008 & 0.009 & 0.009 & 0.028 \\ 
$\rm{[pp][X]}$ (H) & $b\bar{b}$  & 0.11 & 0.13 & 0.12 & 0.38 \\ 
\hline
$\rm{[p][pX]}$ (L) & $b\bar{b}$ & 0.003 & 0.002 & 0.002 & 0.006 \\ 
$\rm{[p][pX]}$ (H) & $b\bar{b}$  & 0.05 & 0.03 & 0.03 & 0.08\\ 
\hline
\hline
Total bgrd (L)  & & 0.17 & 0.17 & 0.18 & 0.54 \\
\hline
Total bgrd (H) & & 0.81 & 0.96 & 0.81 & 2.48 \\
\hline
\end{tabular}
\caption{The final cross sections for the $H\rightarrow b\bar{b}$ and relevant backgrounds for cone and $K_T$ algorithms. The cross sections are defined after a mass window, given by equation \ref{masswin}, has been applied around the Higgs peak. The overlap backgrounds are defined at both low (L) and high (H) luminosity. [pp][X] and [p][pX] jj backgrounds are negligible and not shown. \label{crosssectionfinal}}
\end{table}

If the cross section for central exclusive SM Higgs production in the b-jet channel is $\sim2.5$~fb, then after experimental cuts one would expect around 0.19 events / fb$^{-1}$ if the events could be triggered at L1 and the detectors could be operated at the minimum possible distance from the beam. This corresponds to 5.6 signal events after three years of data acquisition at low luminosity with a background of 21 events. At high luminosity, one would expect a total of 56 signal events with a background of 1032 events after 3 years data taking. Approximately 80\% of the background events at high luminosity are from the overlap backgrounds, and it is these backgrounds that would render the signal unobservable with a reasonable significance, even if a trigger strategy could be developed to retain a high fraction of the events. With the standard experimental configuration, therefore, observation of the SM Higgs in the $b \bar b$ channel would not be possible. 

As discussed above, it might be expected that the asymmetric events could be saved with a high efficiency by triggering on 220m detectors at L1 and reducing the rate at level 2 using 420m detector hits and timing information. Improving the timing resolution to better than 10ps, or adding central detector information into the timing system could significantly improve the rejection of the overlap backgrounds as discussed in section \ref{fasttiming}. If these improvements are made, a significance of approximately  $2$ could be achieved.  Alternatively and additionally, if improvements in trigger efficiency for symmetric tags, b-tagging efficiency and acceptance of the forward detectors could be achieved, or the cross section turns out to be at the upper-end of theoretical expectations, it may be possible to achieve a $3 \sigma$ observation. Since there may well be no other way of measuring the SM Higgs b-quark coupling at the LHC, these possibilities should not be dismissed. We conclude, however, that observing the CEP of the SM Higgs in the $b \bar b$ decay channel will be extremely challenging. We now move on to consider in more detail a scenario in which the CEP cross section is predicted to be $\sim 8$ times higher than the SM prediction.   

\subsection{MSSM Higgs bosons}\label{discussion}

In this section, we perform a similar analysis for Higgs bosons produced for the benchmark $m_h^{\rm{max}}$ scenario as described in section \ref{scenarios}. The cross section for the CEP of the lightest Higgs boson in this scenario is enhanced by a factor of $8$ relative to the Standard Model. 

After the same cuts as for the SM Higgs analysis, except for applying a wider mass window of $\pm 5.2$ GeV for the symmetric events and $\pm 6.7$ GeV for the asymmetric events because of the increased width of the Higgs (3.3 GeV), the signal cross section is 0.54 fb for the 420m + 420m case, and 0.28 (0.85) fb for the 420m + 220m case if the detectors are 5mm/2mm (3mm/1.5mm) from the beam. This would lead to over 150 signal events for the symmetric analysis, and over 200 events for the combined analysis for an integrated luminosity of 300 fb$^{-1}$, again assuming a 100 \% trigger efficiency. Because of the larger mass window, the continuum background within the window also increases by a factor of approximately $2$-$2.5$.  

\subsubsection{Potential for observation of MSSM Higgs Bosons and mass determination}
\label{MSSM}

In order to determine the significance of the MSSM Higgs signal, we construct an event sample consisting of the appropriate number of events after 3 years of data taking from the signal and background Monte Carlos after the cuts described in section \ref{smresults}. We assume several different trigger strategies as described in section \ref{trigger}. We present results for data acquisition at low and high luminosity, i.e. 30~fb$^{-1}$ of data at a luminosity of 10$^{33}$~cm$^{-2}$~s$^{-1}$ and 300~fb$^{-1}$ of data at 10$^{34}$~cm$^{-2}$~s$^{-1}$. In real experimental conditions the data will be taken at many different instantaneous luminosities. 

We perform a fit to the pseudo-data with two curves; one assuming that there is a signal in the region $100 \leq M \leq 140$ GeV and one assuming that there is no signal. The significance, $S$, is then given by 
\begin{equation}\label{signifcalc}
S = \left(\Delta \chi^2 \right)^{\frac{1}{2}}
\end{equation}
where $\Delta \chi^2$  is the difference of the $\chi^2$ of the fits for signal and null hypotheses. We fit the  peak region using a gaussian convoluted with a Lorentzian function; the gaussian represents the mass resolution of the detectors, which is a known parameter. The width of the Higgs boson is left as a free parameter in the fit. We assume that the shape of the background will be known, as it can be measured with high statistics with the forward detectors. The normalisation of the background is left as a free parameter in the fit \footnote{As a cross-check, the fits were repeated using a simple quadratic fit to the background, and similar results were obtained.}.      
 We repeat this method for 500 pseudo-data sets to obtain the average significance of the fits.

\begin{figure}
\centering
\mbox{
	\subfigure[]{\includegraphics[width=.5\textwidth]{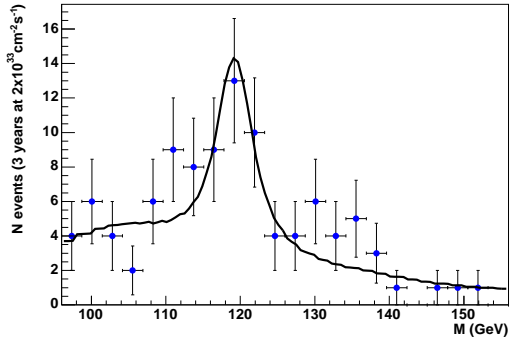}}}
	
\caption{A typical mass fit for 3 years of data taking at $2~\times~10^{33}~$cm$^{-2}$~s$^{-1}$ (60 fb$^{-1}$). The significance of the fit is $3.5 \sigma$ and uses only events with both protons tagged at 420m. The L1 trigger strategy and analysis cuts are described in the text. \label{mssmxsll}}
\end{figure}
\begin{figure}
\centering
\mbox{
	\subfigure[]{\includegraphics[width=.5\textwidth]{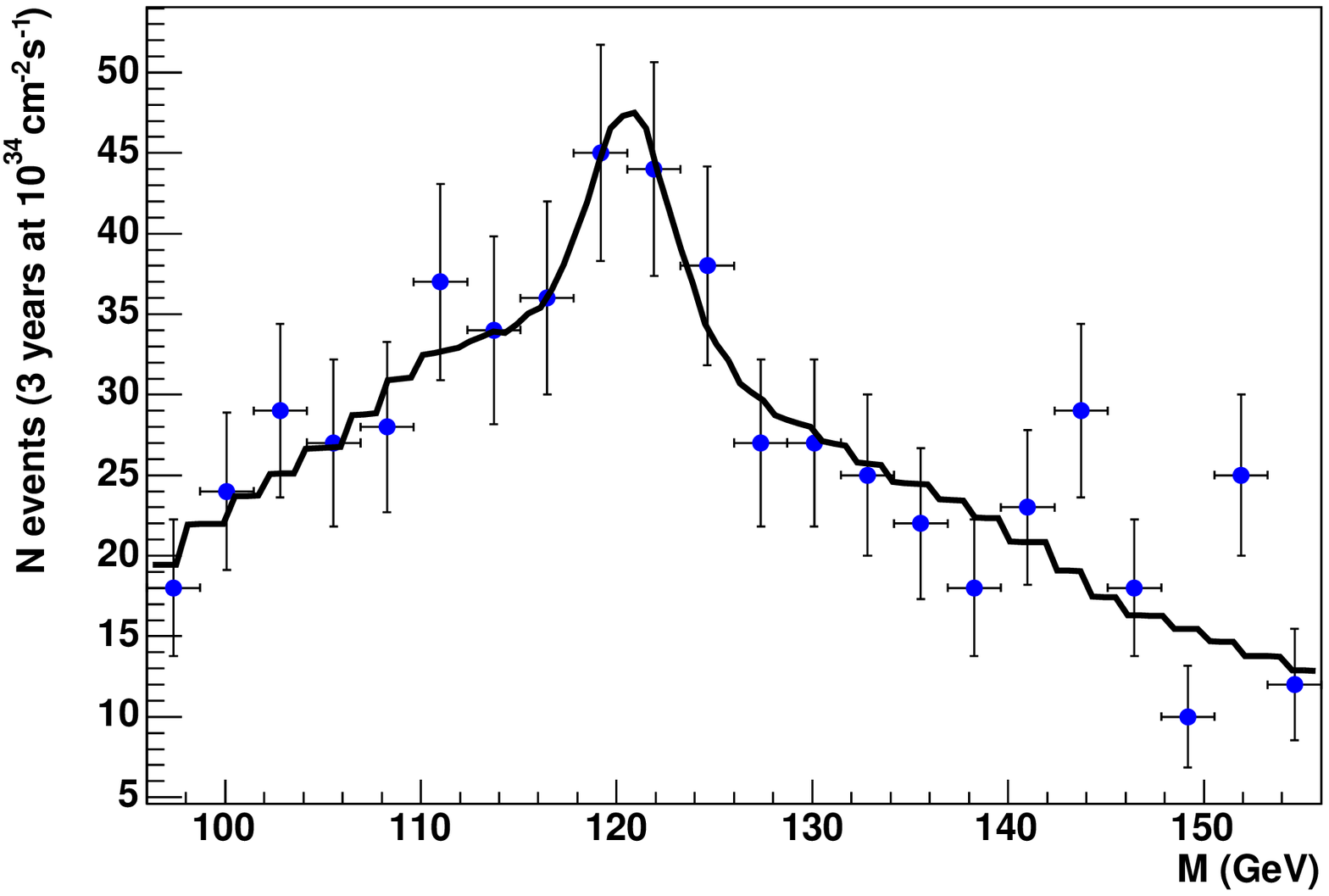}}
	\subfigure[]{\includegraphics[width=.5\textwidth]{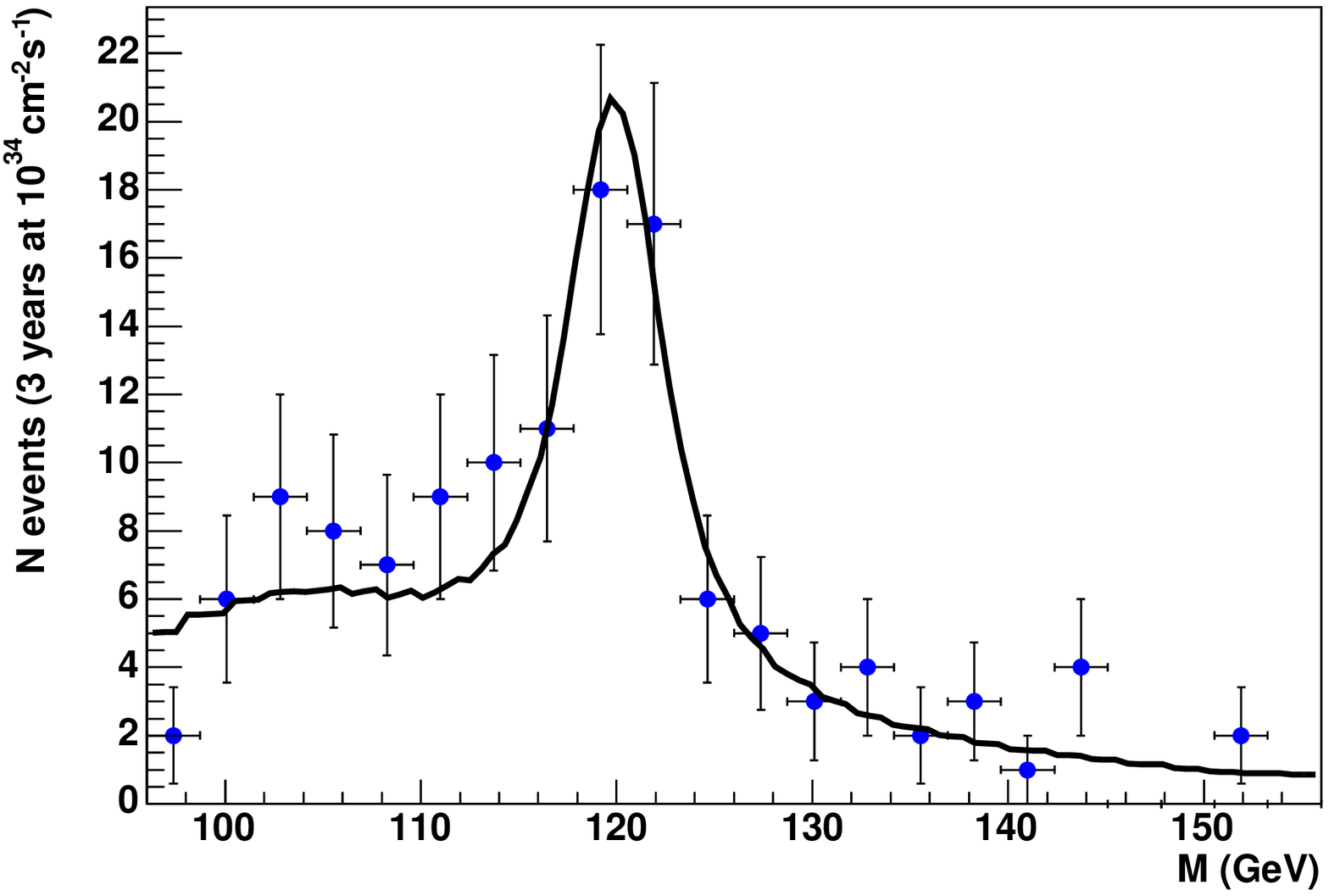}}
}

\caption{ Figure (a) shows a mass fit for 3 years of data taking at $10^{34}~$cm$^{-2}~$s$^{-1}$, for events with both protons tagged at 420m. The significance is $3 \sigma$. The L1 trigger and analysis cuts are described in the text. Figure (c) shows the same high-luminosity data set after removing the overlap background contribution completely, which would be possible with improved timing detectors. The significance is $5 \sigma$ \label{mssmxshl}}
\end{figure}

Examples of three pseudo-data sets and fits are shown in figures \ref{mssmxsll} and \ref{mssmxshl}.  In figure \ref{mssmxsll} we consider 60 fb$^{-1}$ of data taken at  $2 \times 10^{33}$~cm$^{-2}$~s$^{-1}$ with 420m + 420m tags only. The error bars correspond to $\sqrt N$ statistical errors. The L1 triggers are J25 + MU6 and the rapidity gap trigger  (see section \ref{trigger}). Figure \ref{mssmxshl} (a) is for the same experimental configuration, but for $300$ fb$^{-1}$ of data taken at $10^{34}$~cm$^{-2}$~s$^{-1}$. The number of events increases due to the non-prescaled MU6 trigger, but the signal-to-background ratio decreases because the overlap background increases with instantaneous luminosity. Figures \ref{mssmxsll} and \ref{mssmxshl}  (a) are constructed under the assumption that the 420m detectors have a timing resolution of 10ps.  If the timing resolution is improved or timing information from the central detector can be combined with the forward detector information, then the overlap background can be effectively eliminated at 10$^{34}$~cm$^{-2}$~s$^{-1}$ as described in section \ref{fasttiming}. Figure \ref{mssmxshl} (b) shows the high-luminosity data set without the overlap background. 

\begin{figure}
\centering
\mbox{
	\subfigure[]{\includegraphics[width=.5\textwidth]{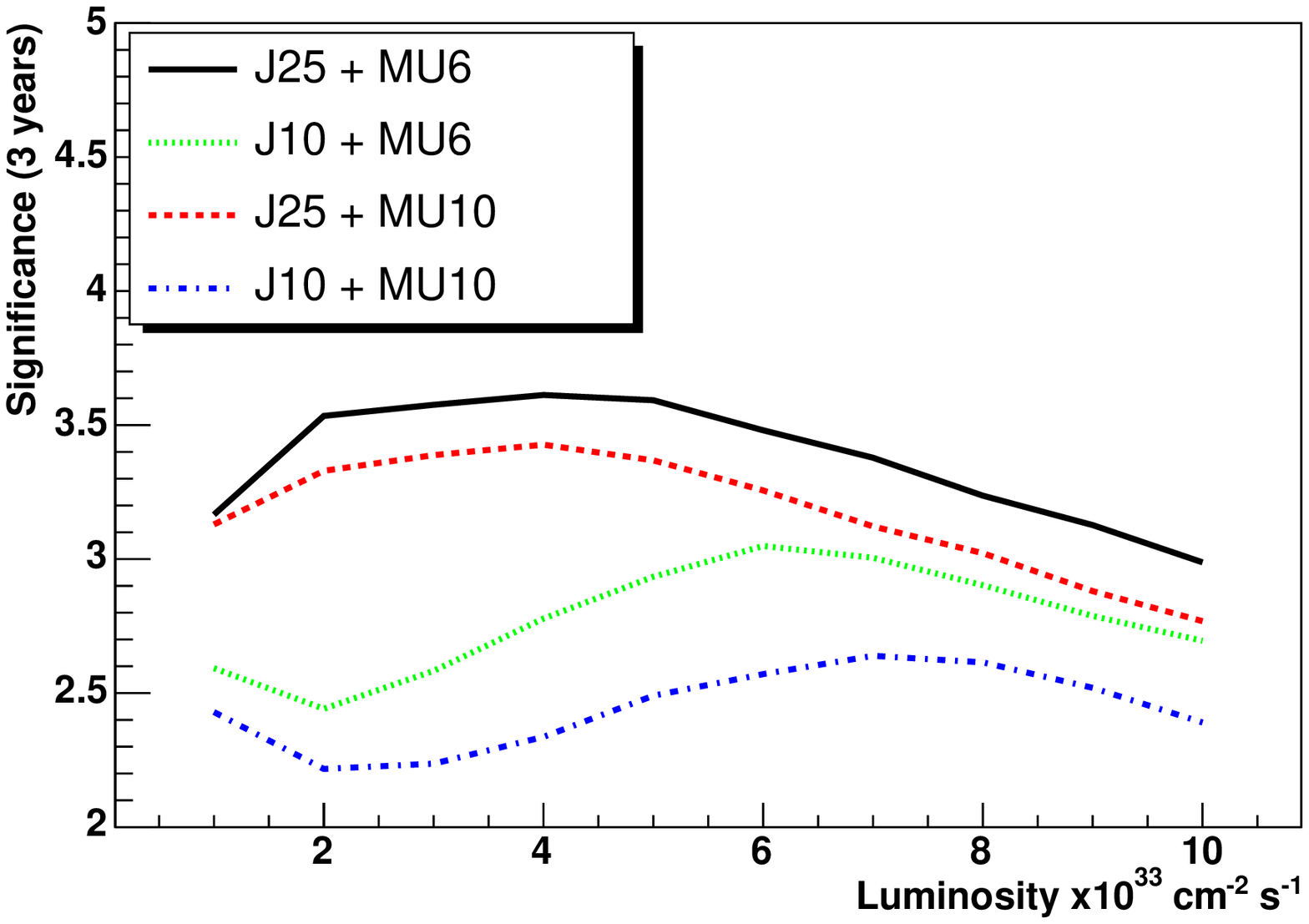}} \quad
	\subfigure[]{\includegraphics[width=.5\textwidth]{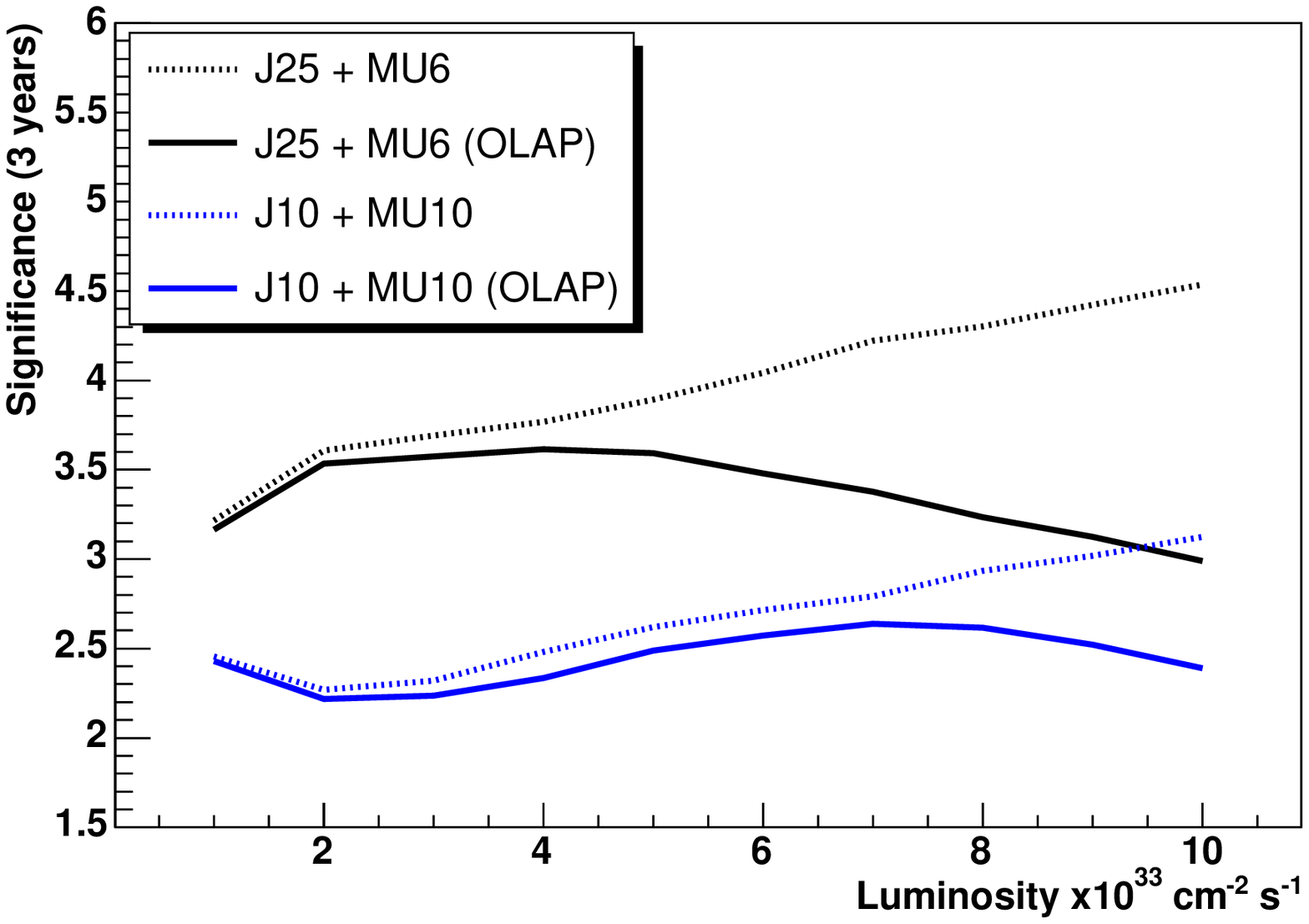}}
	}
\caption{Figure (a) shows the significance of the measurement of the 120 GeV MSSM Higgs Boson for the model described in the text, using the combined muon, rapidity gap and jet rate triggers for the analysis with both protons detected at 420m. Figure (b) shows the effect of completely removing the overlap background on two of the trigger strategies.\label{signifcomb}}
\end{figure}

\begin{figure}
\centering
\mbox{
	\subfigure[]{\includegraphics[width=.5\textwidth]{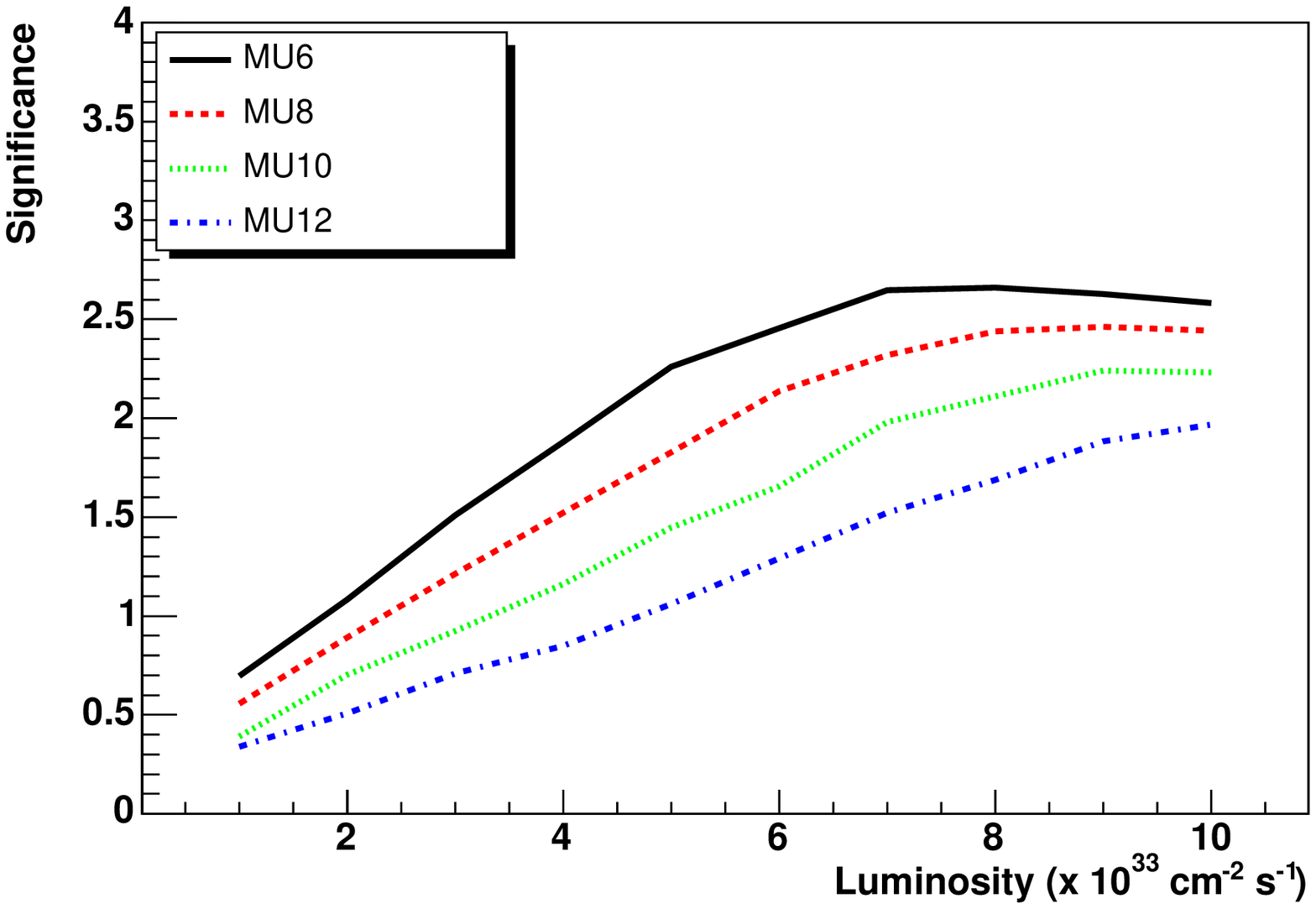}} \quad
	\subfigure[]{\includegraphics[width=.5\textwidth]{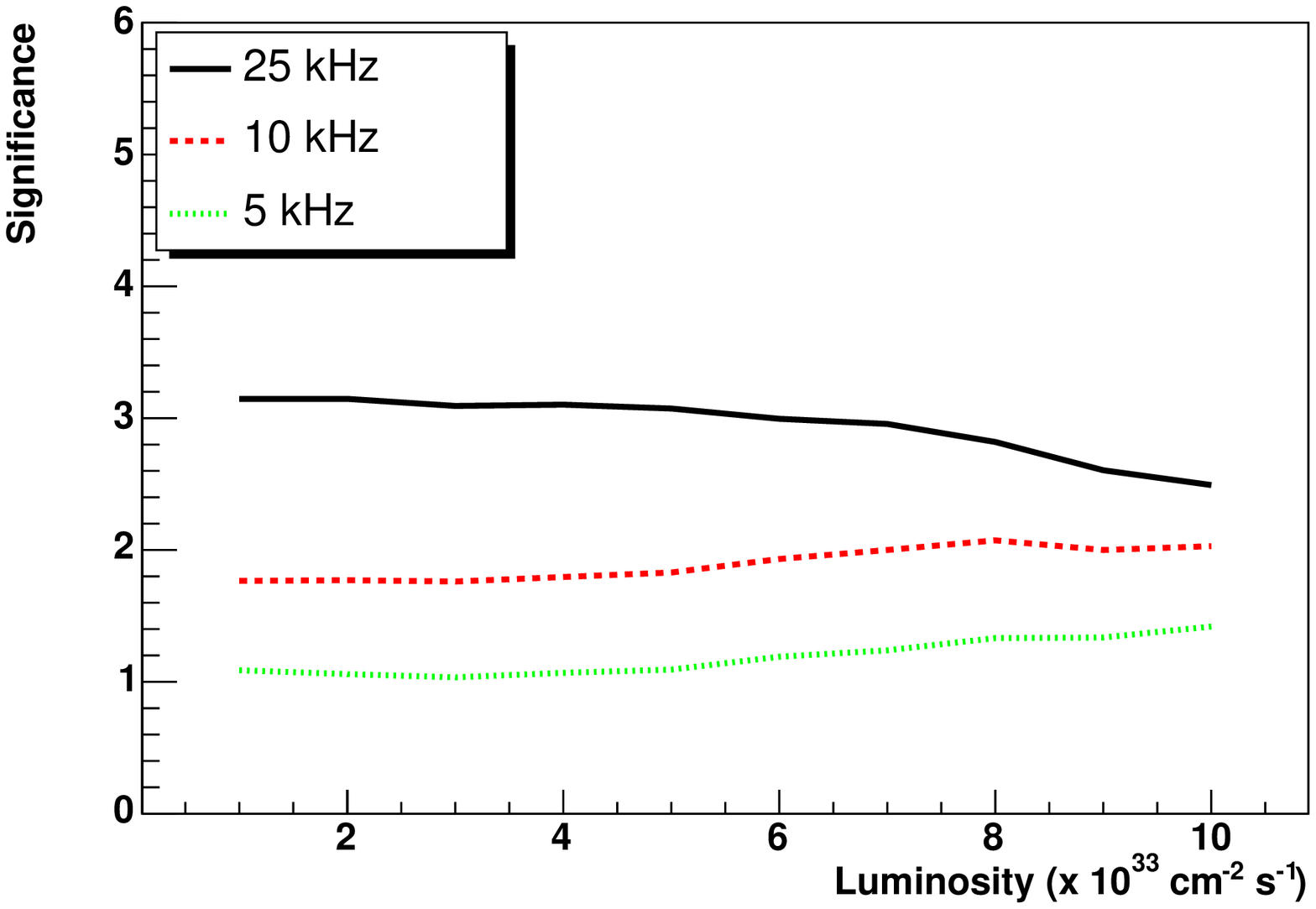}}
	}
\caption{Figure (a) shows the significance assuming that a low transverse momentum muon trigger is used. Figure (b) shows the significance for a pre-scaled jet rate trigger. Both significances correspond to an analysis with both protons tagged at 420m and do not include the contribution from asymmetric proton tagging. \label{signifmuonjets}}
\end{figure}

Figure \ref{signifcomb} (a) shows the significance of the measurement, after 3 years of running at instantaneous luminosities between $10^{33}~$cm$^{-2}~$s$^{-1}$ and $10^{34}~$cm$^{-2}~$s$^{-1}$  assuming that the protons are tagged only at 420m. A significance of above $3$ is achievable in any 3 year period, regardless of the instantaneous luminosity if the high rate jet trigger (J25) is incorporated at ATLAS in conjunction with the MU6 transverse momentum muon trigger. The significance remains approximately constant because although the overlap background increases at high luminosity and the pre-scaled jet trigger J25 retains an approximately fixed number of events, the non-prescaled muon trigger MU6 retains an increasing number of events, and therefore the total number of signal events increases. This effect can also be seen in figures \ref{mssmxsll} and  \ref{mssmxshl}. 
Figure \ref{signifmuonjets} shows the individual contributions of the muon and jet triggers. The rapidity gap trigger contributes very little to the significance at luminosities of $2 \times 10^{34}$~cm$^{-2}$~s$^{-1}$ and above.

Figure \ref{signifcomb} (b) shows the effect of completely removing the overlap background (i.e. using improved time-of-flight rejection). In this case, the more conservative trigger combination of a 10 GeV transverse momentum muon (MU10) and a pre-scaled jet-rate of 10kHz (J10) is enough to achieve a significance of approximately 3 at high luminosity in a 3 year period. For J25 and MU6, the significance approaches 5. These figures show that with improved timing information, running at  10$^{34}$~cm$^{-2}$~s$^{-1}$ is optimal, whereas for the baseline 10ps + 10ps timing with no information from the central detector, data taking at luminosities in the range $2 - 5 \times 10^{33}$~cm$^{-2}$~s$^{-1}$ is optimal.  


\begin{figure}
\centering
\mbox{
	\subfigure[]{\includegraphics[width=.5\textwidth]{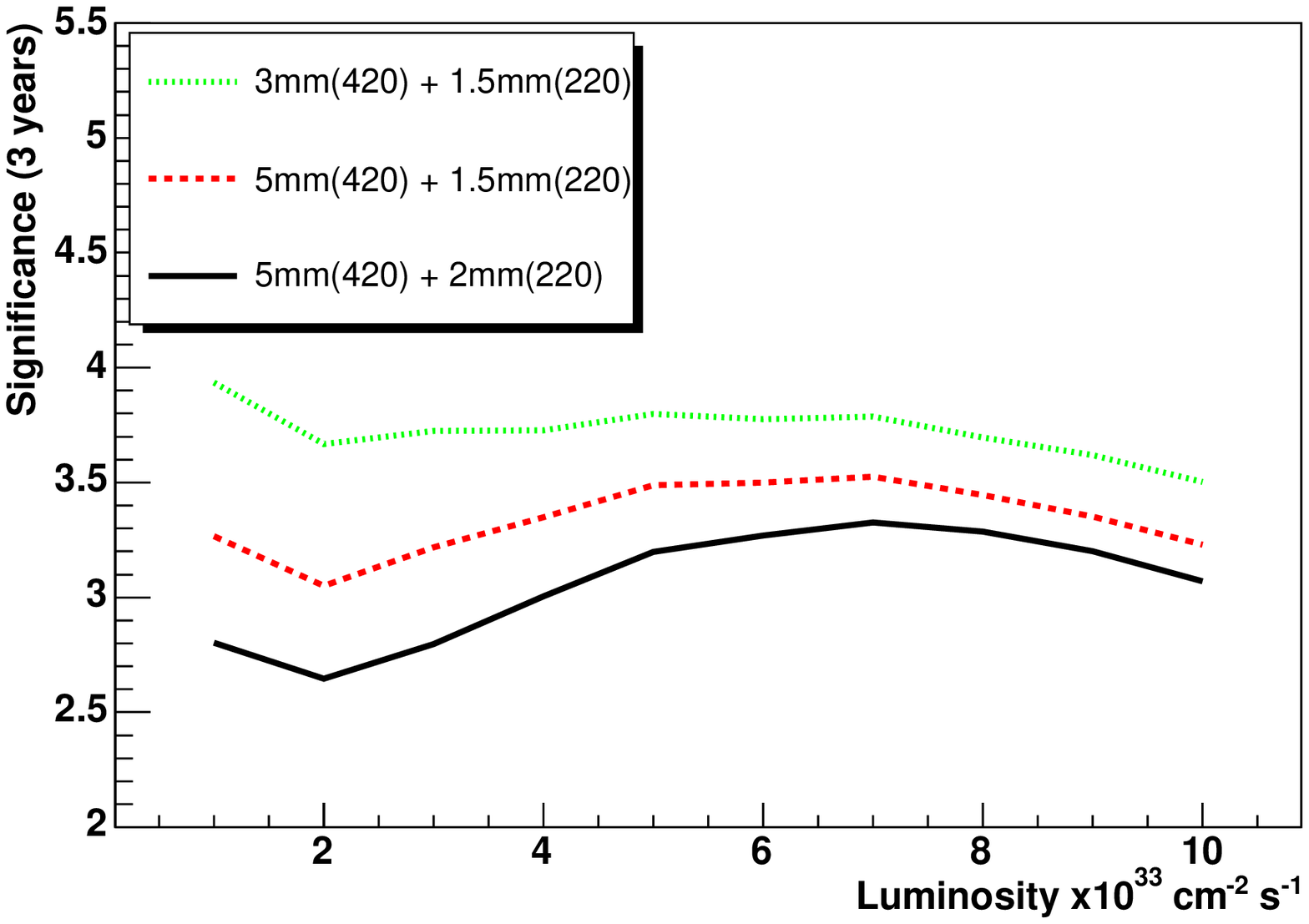}} \quad
	\subfigure[]{\includegraphics[width=.5\textwidth]{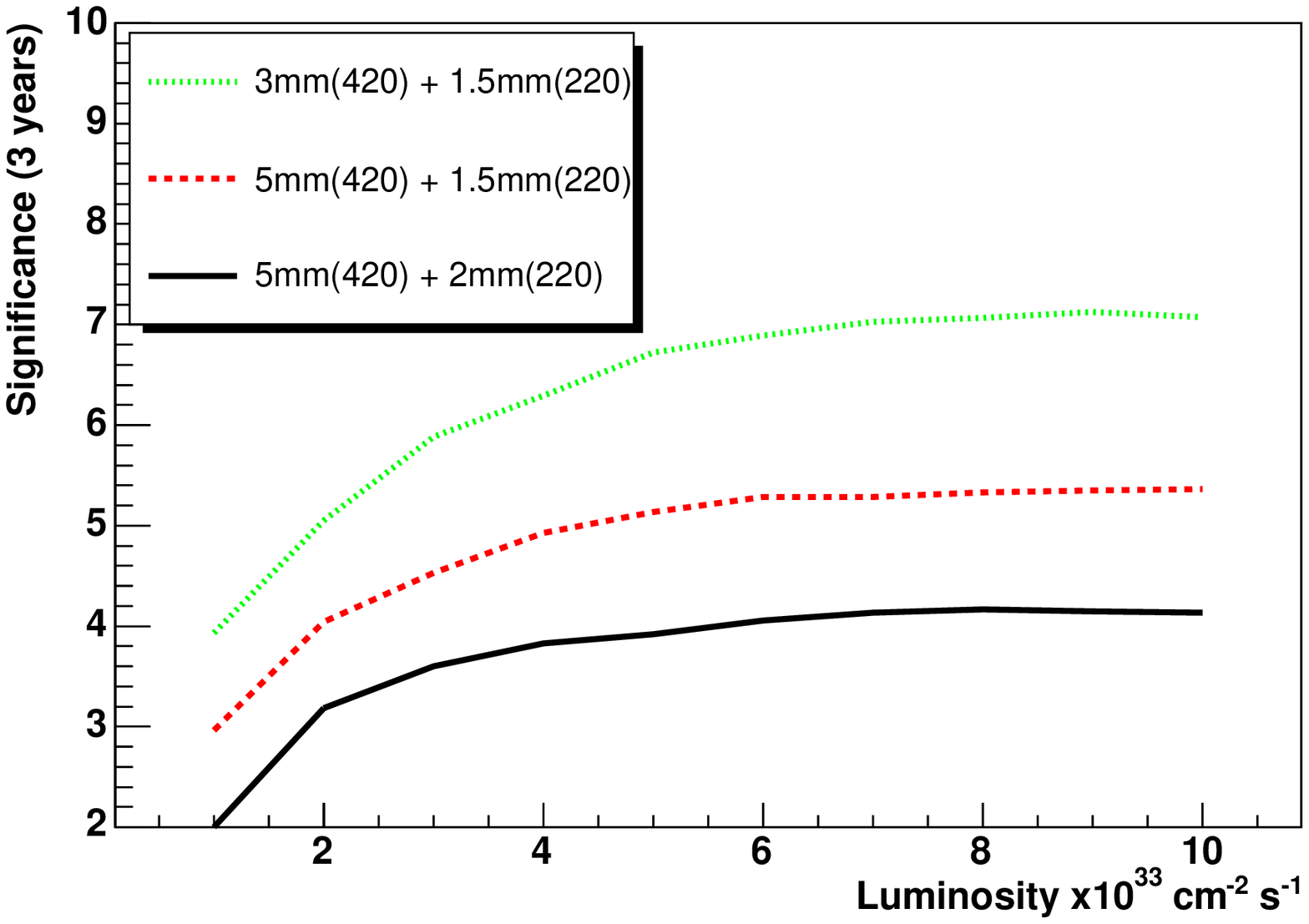}}
	}
\caption{Figure (a) shows the significance of the muon (MU10), rapidity gap and jet rate (J10) triggers for the combined symmetric and asymmetric analyses. This is the more conservative jet rate trigger. Figure (b) shows the contribution from just the asymmetric analyses if all of the events are retained by the level 1 trigger.\label{signifcomb220}}
\end{figure}

Figure \ref{signifcomb220} (a) shows the effect of combining the significance of the analysis using 420m detectors alone with the asymmetric analysis. We use the configuration chosen in section \ref{fp420}, where the detectors at 420m are 5mm from the beam and the detectors at 220m are 2mm from the beam. Here we choose the J10 + MU10 trigger, since this trigger scenario is difficult for 420m detectors alone. The significance rises from $\sim 2.5$ to $\sim 3.0$ for the nominal detector positions (5mm /  2mm)  from the beam. Figure \ref{signifcomb220} (a) also shows the combined significance increases to between 3.5 and 4.0 if the detectors are moved to the closest possible distance from the beam  (3mm /1.5mm). If the overlap background is eliminated then the combined significance rises to between 3.6 and 5.0, depending on the distance of approach to the beam. For the J25 + MU6 trigger (not shown), the significance after 3 years at $10^{34}$~cm$^{-2}$~s$^{-1}$ rises to $3.7 (4.5)$, for detectors at 5mm / 2mm (3mm / 1.5mm) from the beam. If the overlap background is eliminated, the significances are 5.5 (7.1) for the J25 + MU6 trigger.

A further improvement could be made by including the 220m detectors in the level 1 trigger decision \cite{Grothe:2006dj}, as discussed in section \ref{trigger}. 
Figure \ref{signifcomb220} (b) shows the significance of the asymmetric analysis if all of the events are retained by the level 1 trigger on the 220m detector. For the standard 5mm + 2mm detector positions, the asymmetric analysis achieves a significance of 3 at $2\times10^{33}$~cm$^{-2}$~s$^{-1}$. However, at higher luminosities the significance approaches 4. If the detectors are moved to 3mm + 1.5mm, then a significance of more than 5.0 is achieved at a luminosity of $2\times10^{33}$~cm$^{-2}$~s$^{-1}$. If the overlap background is removed, the significance at high luminosity approaches 11 after 3 years of data taking.

\section{Uncertainties on the calculated cross sections}\label{uncertainties}

All the signal and background processes are affected by the uncertainty in the calculation of the soft survival factor. Whilst this is expected to vary to some extent from process to process, to a reasonable approximation it produces a change in the overall normalisation of all cross sections, and is not expected to affect the signal to background ratios quoted in this paper (although it will of course affect the significance). 

In the analysis presented here it is not possible to quantify the effect of pile-up on the jet-finding and the analysis cuts used to define the exclusive sample. Such effects will depend on the performance of the central detector as well as the pile-up conditions. We now deal with the uncertainties specific to each of the signal and background calculations.  
   
\subsection{Signal}
As discussed in section \ref{scenarios}, the cross section for central exclusive production is expected to have at least a factor of two uncertainty. This would also apply to the CEP backgrounds, since the uncertainty enters primarily in the calculation of the luminosity term, which contains the physics of the colour-singlet gluons, and not in the hard sub-process cross section calculation of Higgs, $b \bar b$ or di-gluon production. We note that the CDF collaboration have observed exclusive di-photon candidate events at the predicted rate \cite{:2007na}. It should be noted, however, that the invariant mass of the di-photon system is much lower than 120 GeV, and the uncertainties on the calculation may be significantly larger due to non-perturbative effects. CDF have also presented preliminary results on the search for the CEP of di-jets \cite{Terashi:2007kb}. The data are well described if an exclusive di-jet component is present in the data at the level predicted by ExHuME (KMR). In summary, therefore, the current KMR predictions are consistent with the Tevatron data, but there are uncertainties in extrapolating to LHC energies. 

There are many other parameter choices in the MSSM. For lower values of tan$\beta$, the cross section is not as large, but neither is the width. As $m_A$ is increased, the heavier scalar boson receives an enhanced CEP cross section and the $b\bar{b}$ decay channel has a large branching ratio up to $m_H\sim160$ GeV. We do not discuss these parameter choices and  refer to the results in \cite{Heinemeyer:2007tu}. However, we note that the acceptance of a 220m + 420m forward detector system is approximately constant up to $M \sim 200$ GeV and so the overall conclusions presented in this paper should remain valid for larger $m_A$. For smaller tan$\beta$, the discovery potential will be reduced.

\subsection{Background}

\subsubsection{Double pomeron exchange}

\begin{figure}
\centering

\includegraphics[width=.5\textwidth]{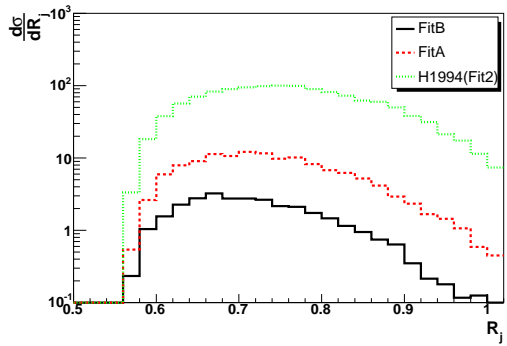}

\caption{The $R_j$ distribution of the DPE $b\bar{b}$ events for three different diffractive PDF fits.\label{dperjfits}}
\end{figure}

 The main source of uncertainty in the DPE processes comes from the diffractive parton distribution functions at high values of $\beta$, the momentum fraction of the pomeron carried by the struck parton. If the value of $\beta$ is large for both partons entering the hard scatter, then the reconstructed $R_j$ value will also be large, and this is the region that contributes to the background. Thus, the final cross section depends strongly on the high-$\beta$ parton distribution of the pomeron. Figure \ref{dperjfits} shows the effect of changing the diffractive pdfs on the $R_j$ distribution of the DPE $b\bar{b}$ events. H1 2006 Fit A and H1 1994 Fit 5 are shown against H1 2006 Fit B, which was chosen as the default fit in this analysis. H1 2006 Fit A increases the final cross section by a factor of approximately 6. Even with such an increase, however, the DPE background stays well below the CEP backgrounds, and the overlap background at high luminosity. If H1 1994 Fit 5 is used there is a large increase in cross section in the large $R_j$ region ($\times70$). The reason for this is that the DGLAP evolution of this diffractive PDF is frozen above $Q^2 = 75$ GeV$^2$, which means that there is an increased probability of large $\beta$ values because there is no radiation between the pdf freeze-out and the hard scale. The 2006 pdfs include data up to $Q^2 = 1600$ GeV$^2$ and are evolved outside of this region, and are therefore likely to give a better estimate of the background. H1 2006 Fit B has a similar high-$\beta$ behavior to the MRW diffractive pdf fits \cite{Khoze:2007hx,Martin:2006td}.   
 
\subsubsection{Overlap background}

The overlap background calculation of equation \ref{olapxs} requires knowledge of the fraction of events that produce one or two forward protons within the acceptance of the forward detectors. Table \ref{fractions} shows that for a single forward proton measured in the 420m region ($0.002 \leq \xi \leq 0.02$), there is a consensus between different models and that the fraction of events that have such a proton is predicted to be $\sim$1\% at the LHC. For larger values of $\xi$, the non-diffractive contribution is not well constrained and this gives an uncertainty on the overlap cross section with one proton detected at 420m and one at 220m. However, the non-diffractive protons typically have larger values of $\xi$ than the diffractive protons and so have a smaller chance of producing a central mass in the $\sim120$ GeV range. In fact, the final cross section for the asymmetric-tagged [p][X][p] background given in table \ref{crosssectionfinal}, which contains a non-diffractive component, is only 30\% larger than without the non-diffractive contribution. If the non-diffractive contribution is uncertain to $\sim50\%$, as indicated in table \ref{fractions}, then the asymmetric-tag overlap backgrounds are uncertain to $\sim 15\%$.

For the [pp][X] events, the uncertainty is larger. In this analysis, we use the total DPE cross section predicted by PHOJET to give the [pp] event fraction at the LHC. However, PHOJET predicts that a large component of the single forward proton spectrum originates from DPE events. A similar result was observed in \cite{Albrow:2006xt}, where it was found that the hard-scattering DPE cross section predicted by PHOJET was much larger than that predicted by POMWIG. This implies that the [pp][X] cross section could be much smaller at the LHC than we assume here. 

The final uncertainty is in the number of charged tracks produced by an inclusive di-jet overlap event at the LHC. This will affect the efficiency of the charged track multiplicity cuts on the [pp][X] and [p][X][p] backgrounds. The number of charged tracks outside of the di-jet system will be dominated by initial state radiation and multi-parton interactions. In this analysis we use HERWIG with JIMMY tuned to CDF data (tune A in \cite{jimmytune}). A second tune in \cite{jimmytune} (tune B) is also able to fit the data. Pythia may be used with a variety of different tunes \cite{Field:2006gq}, all of which fit the CDF data. The tunes each predict a different number of charged tacks in the transverse region when extrapolated to LHC energies. Table \ref{tunes} shows the percentage of the inclusive di-jet backgrounds that pass the $N_C$ and $N_C^{\perp}$ cuts with the different available tunes. Apart from the Pythia DW tune, the rejection factors vary by approximately a factor of 2.5. However, it should be noted that the ATLAS detector is capable (although with a reduced efficiency)  of measuring charged particles that have $p_T < 0.5$ GeV and this could increase the efficiency of the charged multiplicity cuts.

\begin{table}[t]
\centering
\begin{tabular}{|c|c|c|c|}
\hline
Generator & Tune & $N_C^{\perp} \leq 1 $ & $N_C^{\perp}\leq1$ and \\
& & & $N_C\leq3$ \\
\hline
HERWIG/JIMMY & A & 3.2 & 1.0\\
HERWIG/JIMMY & B & 4.1 & 1.2  \\
Pythia & DWT & 9.5 &  2.2 \\
Pythia & DW & 14.9 &  3.7 \\
Pythia & ATLAS & 7.3 &  2.3\\
\hline
\end{tabular}
\caption{Percentage of inclusive di-jet backgrounds that pass the charged multiplicity cuts. \label{tunes}}
\end{table}%

\section{Conclusions}
\label{conclusions}
We have shown that the central exclusive production of the lightest scalar Higgs boson, for the choice of MSSM parameter space described in section \ref{scenarios}, can be observed with a significance of at least $3 \sigma$ in the $b\bar{b}$ decay channel within 3 years of data taking at the LHC, if suitable proton tagging detectors are installed around ATLAS and CMS and the current predictions for the CEP cross section are correct. We have evaluated the most important backgrounds and shown that they can be rejected with high efficiency using a set of exclusivity variables. We have also shown that a fraction of b-jet events can be retained by the currently forseen Level 1 trigger hardware, if the trigger strategies we outline are adopted. Whilst we have only considered a particular choice of parameters in detail, the general conclusions should hold for a wide range of scenarios. As a general rule, if the CEP cross section for the production of the Higgs boson is greater than $\sim10$ fb, and the Higgs boson decays predominantly to b-quarks, then the analysis presented here should apply if the decay width is not too large. The analysis will also apply to any new particle that decays predominantly to b-quarks.   

\begin{figure}
\centering
\mbox{
	\subfigure[]{\includegraphics[width=.5\textwidth]{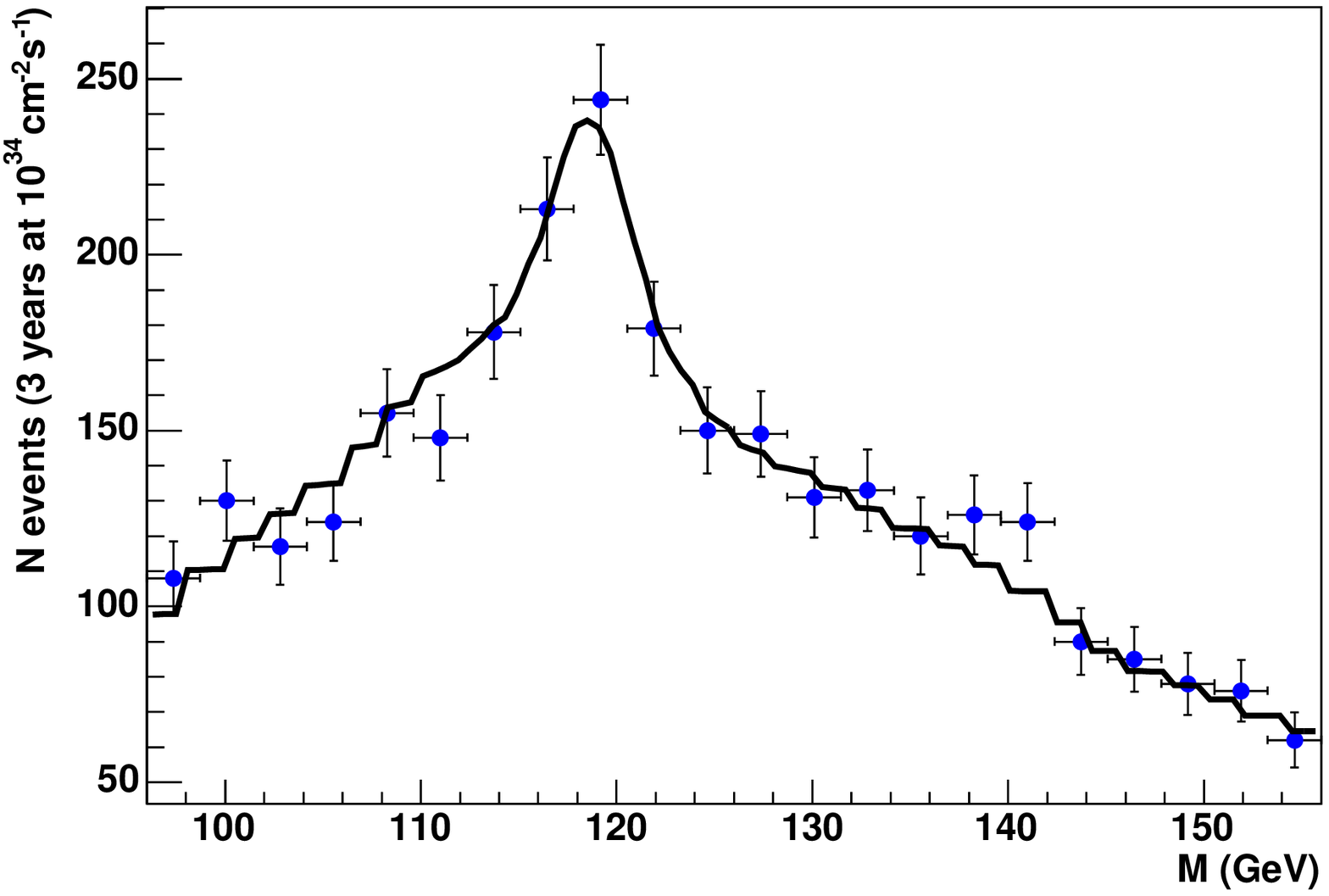}} \qquad
	\subfigure[]{\includegraphics[width=.5\textwidth]{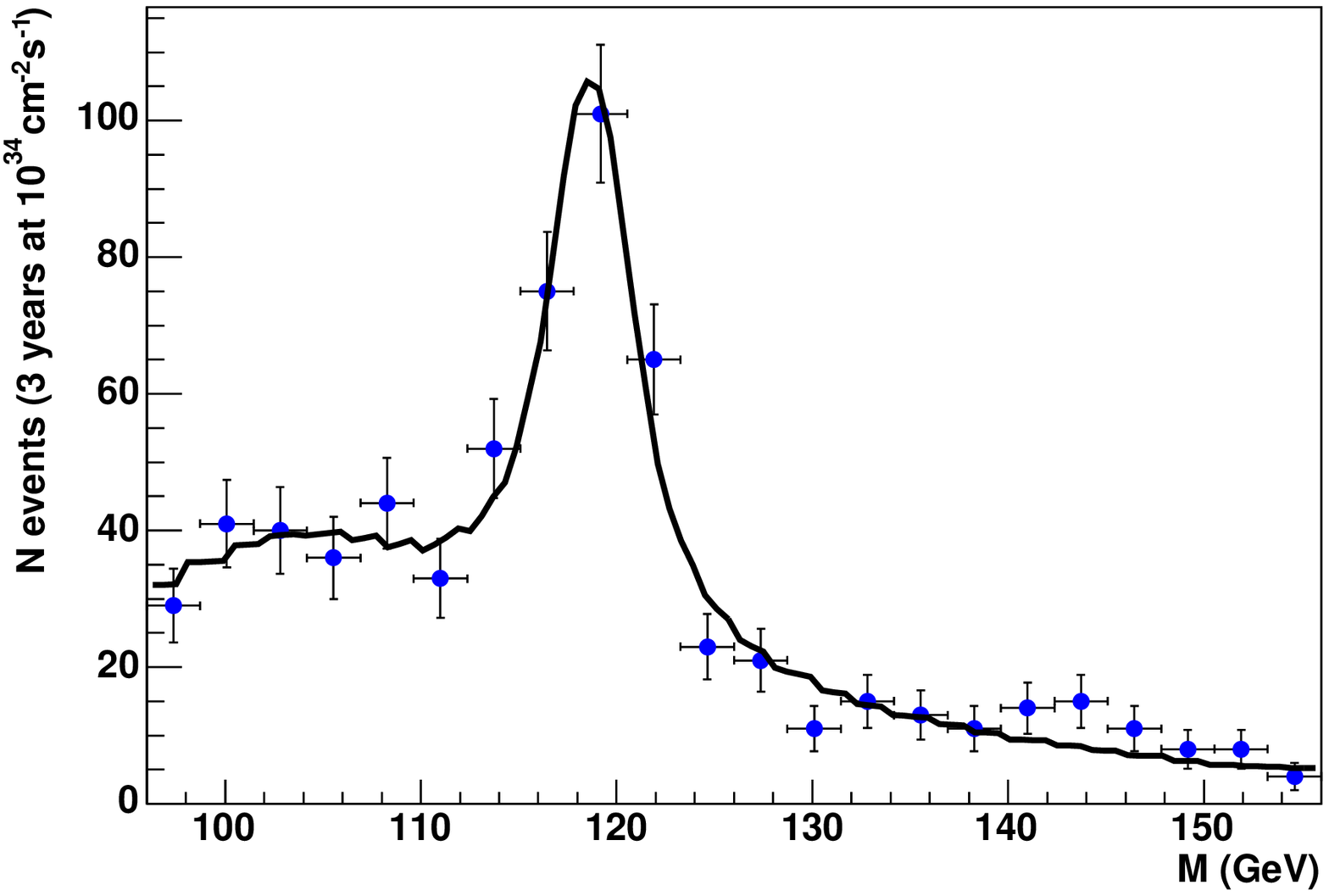}}
	}
\caption{Figure (a) shows a typical mass fit for the analysis at high luminosity if it is assumed that a trigger upgrade allows the 420m detectors to be included in the L1 trigger at ATLAS and CMS. The significance is approximately $6.5 \sigma$. Figure (b) shows the same data after removing the overlap contribution. The significance is approximately $10 \sigma$. \label{upgrade}}
\end{figure}

It is worth speculating what the future experimental strategy might be if a Higgs sector such as the $m_h^{max}$ scenario of the MSSM is discovered at the LHC or Tevatron, and the CEP channel proves to be observable with the forward detector configurations currently proposed. The largest loss of CEP signal events comes from the limited acceptance of the proton detectors (between $25\%$ and $40\%$ depending on detector configuration) and the L1 trigger efficiency, which is at best around $20 \%$ at high luminosity for the strategies we consider in this paper. If a hardware upgrade of the L1 trigger systems of ATLAS and CMS were to increase the trigger latency such that the 420m detector signals could be included, then a trigger efficiency of close to $100 \%$ could be achieved. Figure \ref{upgrade} (a) shows a typical mass fit, with 300 fb$^{-1}$ of data taken at high luminosity using 420m detectors alone (i.e. symmetric events), assuming 100\% trigger efficiency. The significance of this signal is approximately $6.5 \sigma$. If it is assumed that, in addition, fast timing detector improvements can be made as described in section \ref{MSSM}, then the significance rises to nearly $10 \sigma$, as shown in figure \ref{upgrade} (b). In this case, a measurement could be made within 100 fb$^{-1}$. Combining the symmetric and asymmetric analyses increases the significance such that a 5$\sigma$ measurement could be achieved within 30 fb$^{-1}$. The mass of the lightest scalar Higgs can be measured with an accuracy of better than $1$ GeV. For the $m_h^{max}$ scenario considered here, the width of the lightest Higgs is not much larger than the mass resolution of the proton detectors, and therefore the extraction of the Higgs width from the fits is marginal. In other scenarios with widths in excess of $\sim 5$ GeV, however, a direct measurement of the width would also be possible. We remind the reader that observation of any particle in the CEP channel would provide a direct measurement of the quantum numbers of the particle. Furthermore, if the pseudo-scalar Higgs is close in mass to the lightest scalar Higgs, then CEP would provide an unambiguous separation of the two states since the pseudo-scalar cannot be produced. Finally, with such a trigger strategy in addition to improved fast timing to reduce the overlap backgrounds, the SM Higgs may be obervable in the b-jet channel. 

\section*{Acknowledgments}

We would like to thank Mike Albrow, Michele Arneodo, Andrew Brandt, Albert DeRoeck, Jeff Forshaw, Valery Khoze, Henri Kowalski, Paul Newman, Will Plano, Misha Ryskin, Marek Tasevsky and Chris Tevlin for interesting discussions and suggestions throughout this project. This work was funded in the UK by STFC and The Royal Society.

\end{document}